\documentclass[prd,twocolumn,amsmath,amssymb,floatfix,superscriptaddress]{revtex4-1}

\usepackage{graphicx}
\usepackage{amssymb}
\usepackage{amsmath}
\usepackage{hyperref}
\usepackage{color}

\DeclareMathAlphabet\mathbfcal{OMS}{cmsy}{b}{n}
\definecolor{darkgreen}{cmyk}{0.85,0.2,1.00,0.35} 
\definecolor{purple}{cmyk}{0.5,1.0,0,0}

\def\barray{\begin{array}}
\def\earray{\end{array}}
\def\be{\begin{equation}}
\def\ee{\end{equation}}
\def\ben{\begin{equation} \nonumber}
\def\een{\end{equation}}
\def\ban{\begin{eqnarray*}}
\def\ean{\end{eqnarray*}}
\def\ba{\begin{eqnarray}}
\def\ea{\end{eqnarray}}

\def\({\left(}
\def\){\right)}

\begin{document}

\title{Transition radiation at radio frequencies from ultra-high energy neutrino-induced
showers}
\author{Pavel Motloch}
\affiliation{Kavli Institute for Cosmological Physics \& Department of Physics, University of Chicago, Chicago, IL
60637, USA}
\author{Jaime Alvarez-Mu\~niz}
\affiliation{Departamento de F\'isica de Part\'iculas \& Instituto Galego de F\'isica de
Altas Enerx\'ias, Universidade de Santiago de Compostela, 15782 Santiago de Compostela,
Spain}
\author{Paolo Privitera}
\affiliation{Kavli Institute for Cosmological Physics \& Department of Physics, University of Chicago, Chicago, IL
60637, USA}
\author{Enrique Zas}
\affiliation{Departamento de F\'isica de Part\'iculas \& Instituto Galego de F\'isica de
Altas Enerx\'ias, Universidade de Santiago de Compostela, 15782 Santiago de Compostela,
Spain}

\begin{abstract}
Coherent radiation at radio frequencies from high-energy showers fully contained in a
dense radio-transparent medium - like ice, salt, soil or regolith - has been extensively
investigated as a promising technique to search for ultra-high energy (UHE) neutrinos.
Additional emission in the form of transition radiation may occur when a neutrino-induced
shower produced close to the Earth surface emerges from the ground into atmospheric air.
We present the first detailed evaluation of transition radiation from high-energy showers
crossing the boundary between two different media.  We found that transition radiation is
sizable over a wide solid angle and coherent up to $\sim$ 1 GHz. These properties
encourage further work to evaluate the potential of a large-aperture UHE neutrino
experiment based on detection of transition radiation.  
\end{abstract}

\maketitle
\section{Introduction}
\label{sec:introduction}

The nature and origin of UHE cosmic rays (UHECRs) with energies around $10^{20}$ eV is one
of the most puzzling questions in particle astrophysics \cite{Nagano:2000ve,Watson:2013cla}.  Charged
cosmic rays with energy less than $\sim$~10 EeV are significantly scattered by the
galactic magnetic field and cannot be traced back to their origin.  At higher energies
where deflections become less important, the cosmic ray flux falls steeply
\cite{Abraham:2010mj,Fukushima:2015bza}, requiring giant experiments like the Pierre Auger
Observatory \cite{Aab:2015zoa} and the Telescope Array \cite{AbuZayyad:2012kk} to collect
reasonable statistics. Neutrinos, which travel undeflected through the universe, may
provide important clues on the origin of UHECRs \cite{Kotera:2011cp}. In fact, cosmic rays
interacting with matter and radiation present at the accelerating source, or encountered
along their path towards Earth, may produce UHE neutrinos. Recently, extraterrestrial
neutrinos of energies up to few PeV have been detected by the IceCube Neutrino Observatory
\cite{Aartsen:2013jdh}. For neutrino energies in the EeV range, a promising detection
technique is based on the radio signal from the neutrino-induced shower in a dense
radio-transparent medium. Ice has been proven to be an ideal medium for this technique by
pioneering experiments such as RICE \cite{Kravchenko:2001id}, which buried radio antennas
in the South Pole ice, and ANITA \cite{Gorham:2008dv}, which looks with balloon-flown
radio antennas over a large portion of the Antarctica ice cap. Large ground arrays of
antennas, buried in ice (ARA \cite{Allison:2011wk}) or on the ice surface (ARIANNA
\cite{Barwick:2014pca} and GNO \cite{Wissel:2015aa}) are now under development. In another
interesting approach, several groups \cite{James:2009sf,Gorham:2003da,
Beresnyak:2005yh} have pointed radiotelescopes towards the Moon, searching for
radiopulses from neutrino interactions in the lunar regolith.  All of these experiments
are based on the Askaryan effect \cite{Askaryan:1962oya} - the emission of coherent
Cherenkov radiation from the $\sim20\%$ excess of electrons present in an electromagnetic
shower.

In this paper we explore transition radiation (TR) \cite{Ginzburg:1945zz} at radio
frequencies as a possible way to detect UHE neutrinos. Transition radiation is emitted by
charged particles when crossing the boundary between two media with different indices of
refraction. From the theoretical viewpoint, TR can be understood as the radiation required
to match at the boundary the electric field produced by the charged
particle in the bulk of the two media. It has been well studied both theoretically and
experimentally; for a comprehensive review of transition radiation, see for example
\cite{Ter-Mikaelian:1972aa}. 
However, there are only a few estimates in the literature of TR from cosmic ray showers
- either crossing from atmospheric air to a denser medium (e.g. clouds
\cite{Gazazian:2001aa} or ground \cite{deVries:2015oda})
or emerging from ground into air (a rough estimate is given in \cite{Falcke:2004aw}).
While the methods presented in this paper
are of general application, we are mostly motivated by the study of TR emitted when a
shower, originated by a UHE neutrino interaction below the Earth surface, escapes the
ground into the atmosphere. In fact, the net electric charge corresponding to the excess
electrons of the shower will produce an upward-going transition radiation when passing
through the ground - atmosphere boundary. Since showers in dense media are rather compact
in size, we expect the TR to be coherent up to GHz frequencies, with significant
advantages in the cost of detectors and background noise level. In this work we present
the first detailed evaluation of transition radiation from high-energy showers crossing
the boundary between two different media.

In Section \ref{sec:calculation}, we present a general method to calculate transition
radiation from a shower crossing the boundary between two media using the far-field limit.
This method extends the well-known Zas-Halzen-Stanev (ZHS) algorithm \cite{Zas:1991jv} to
account for boundary matching of the radiation field.  In ZHS, the trajectory of each
particle in a homogeneous medium is approximated by linear segments, sufficiently small
that the particle can be assumed to move uniformly, and the corresponding electromagnetic
radiation is computed exactly in the Fraunhofer limit from the Maxwell
equations~\cite{Zas:1991jv}.  In the extended ZHS algorithm, hereafter called ZHS-TR, a
particle segment crossing the two media is split at the boundary \cite{James:2010vm} and
reflection and refraction of the radiation emitted by the two track segments is then
included in a standard way~\cite{Ginzburg:1990aa}. 

In Section \ref{sec:results}, the properties of TR emitted by a high energy shower when
crossing from a dense medium to air are studied in detail. In particular, we investigate
the spectral characteristics of the TR signal and their dependence on the shower energy
and zenith angle, on the stage of the shower development at the boundary, on the energy of
the shower particles, and on the type of dense medium.  

Conclusions are drawn in Section \ref{sec:discussion}.

\section{Transition radiation from high-energy showers: calculations}
\label{sec:calculation}
\subsection{The ZHS Monte Carlo}
\label{subsec:ZHS}

Our calculation of radio emission from particle showers crossing the boundary between two
media is based on the methods \cite{Allan:1967,Zas:1991jv, GarciaFernandez:2012yz}
implemented in the ZHS Monte Carlo, which simulates electromagnetic showers and their
associated coherent radio emission up to EeV energies \cite{AlvarezMuniz:2009zzc}.
Originally developed for the Fraunhofer limit in homogeneous ice \cite{Halzen:1990vt}, it
has been extended to other dielectric homogeneous media \cite{AlvarezMuniz:2009zzc,
AlvarezMuniz:2005qb} and to reproduce near field effects by dividing the particle
trajectories in sufficiently small sub-tracks~\cite{AlvarezMuniz:2000fw}.  To provide the
reader with the necessary background, the methods currently implemented in the ZHS Monte
Carlo are reviewed in the following.

The ZHS Monte Carlo follows electrons and positrons interactions down to a kinetic energy
threshold of 100 keV. Charged particles below this threshold contribute negligibly to the
electric field, which is proportional to the particle tracklength \cite{Zas:1991jv}.  The
simulation includes bremsstrahlung, pair production, and the interactions responsible for
the generation of the excess charge (M\o ller, Bhabha, Compton scattering and
electron-positron annihilation). In addition, multiple elastic scattering (according to
Moli\`ere's theory) and continuous ionization losses are implemented.  The track segment
between two consecutive particle interactions is divided into multiple linear segments,
along which the particle is assumed to move with constant velocity. To ensure a proper
calculation in the Fraunhofer limit, the maximum equivalent depth of each sub-track does
not exceed 0.1 radiation lengths. Also, for low energy particles the size of the sub-track
is required to be smaller than the particle range, and this step size is used to evaluate
ionization losses and multiple elastic scattering. Accurate bookkeeping of the absolute
timing is maintained during particle propagation, including geometrical time delays as
well as those introduced by different particle velocities (a uniform energy loss along the
sub-track is assumed for this purpose). An approximate account is also made of the time
delay induced by multiple elastic scattering.
From the position and time of the endpoints of the sub-tracks, the frequency spectrum of
the electric field is derived. The total electric field at the observer's location is then
calculated by superposition of the electric field from each sub-track, with relative phase
shifts due to different starting point positions and time delays properly taken into
account. 

The electric field $\boldsymbol{E}(t, \boldsymbol{x})$ produced by a charged particle
moving with uniform velocity $\boldsymbol{v}$ between two fixed points in a homogeneous
medium can be derived from Maxwell equations, and has a frequency Fourier transform
\footnote{Note we use a non-standard convention.}
\be
	\boldsymbol{E}(\omega, \boldsymbol{x}) = 2 \int_{-\infty}^\infty \boldsymbol{E}(t,
	\boldsymbol{x}) e^{i \omega t} dt 
\ee
given by \cite{Zas:1991jv}:
\be
	\label{ZHSFormula}
	\boldsymbol{E}(\omega, \boldsymbol{x}) = \frac{e \mu_r}{2 \pi \epsilon_0 c^2} i \omega \frac{e^{i k R}}{R}
	e^{i(\omega - \boldsymbol{k} \cdot \boldsymbol{v})t_1} \boldsymbol{v}_\perp
	\left[\frac{e^{i(\omega - \boldsymbol{k} \cdot \boldsymbol{v}) \delta t} -
	1}{i(\omega - \boldsymbol{k} \cdot \boldsymbol{v})}\right] .
\ee
In Eq.~\eqref{ZHSFormula}, $R$ is the distance from the track to the observer, which is
assumed to be large enough for the Fraunhofer regime to be valid.  Wave vector
$\boldsymbol{k}$ points in the direction from the track to the observer and has modulus
${k} = \omega/c'$, with $c'$ the speed of light in the medium. Also,
${\boldsymbol{v}}_\perp = - \hat{ \boldsymbol{k}} \times (\hat{\boldsymbol{k}} \times
\boldsymbol{v})$ is the component of $\boldsymbol{v}$ in the                    plane
perpendicular to the unit vector $\hat{\boldsymbol{k}} = \boldsymbol{k}/|\boldsymbol{k}|$.
Finally, $t_1$ and $t_2 = t_1 + \delta t$ are the absolute times at which the particle
passes through the starting point and endpoint of the sub-track, respectively.  In
magnetic materials, $\mu_r$ is the relative permeability of the medium.

\subsection{The ZHS-TR algorithm}
\label{sec:algorithm}

\subsubsection{General case: Transition between two media}
\label{B1}
The original ZHS Monte Carlo described in Sect.~\ref{subsec:ZHS} has been mostly used in
simulations where the shower is contained in a single dense medium, which also hosts the
observer. Several modifications are required to simulate a shower extending over two
different media.  In the following, we will detail the ZHS-TR algorithm for a shower
starting in a medium of refractive index $n_1$, separated by a planar boundary from a
second medium of refractive index $n_2$ where the observer is located. This configuration
accounts for the case of a shower initiated below the Earth surface by an UHE neutrino
interaction, and emerging in the atmosphere where it is detected. However, the algorithm
is general and can be used for other cases (for example, the shower could start in air, or
the observer could be located in medium 1).

\begin{figure}
\center
\includegraphics[width = 0.48 \textwidth]{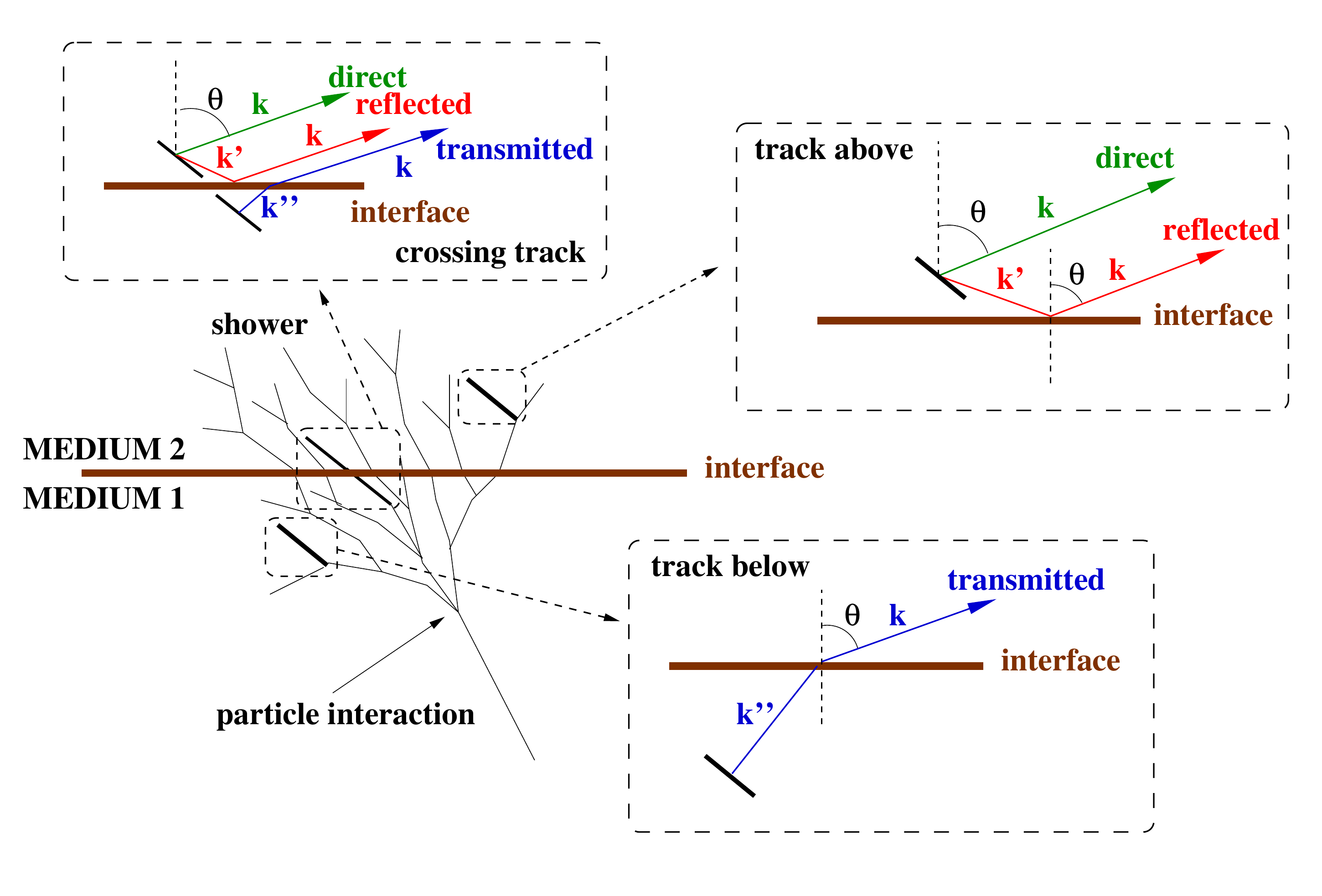}
\caption{Schematic illustration of the ZHS-TR algorithm (see text for a detailed
description). {The three cases shown correspond to a sub-track crossing the interface, a
sub-track in the same medium as the observer (track above) and a sub-track in the other
medium (track below).}}
\label{fig:sketch_contributions}
\end{figure}

As in the original ZHS algorithm, a particle track is approximated by linear segments
along which the particle is assumed to move with constant velocity. However, the electric
field from a given sub-track {can include one, two or three contributions, depending on
whether the sub-track is fully contained in medium 1, in medium 2 (where the observer is)
or crossing the boundary}. The different
contributions are pictorially represented in Fig.~\ref{fig:sketch_contributions}. We
calculate the electric field at the observer position, with the sub-track viewed in the
direction $\hat{\boldsymbol{k}}$. 
{Also, we assume that Fraunhofer conditions apply, namely $\lambda R \gg D^2$ and
$R\gg D$, where $\lambda$ is the wavelength of the radiation and $D$ the characteristic size of the
sub-track. With this approximation the factor $1/R$ in Eq.~\eqref{ZHSFormula} can be
factored out as if it was constant. }

For tracks fully contained in medium~1, the only contribution comes from the electric
field (Eq.~\eqref{ZHSFormula}) emitted into direction $\hat{\boldsymbol{k}''}$ (the
direction of a ray refracted at the boundary into direction $\hat{\boldsymbol{k}}$
according to Snell's law, see Fig.~\ref{fig:sketch_contributions}). The radiated electric
field is then decomposed into $s$ and $p$ polarizations (perpendicular and parallel to the
plane of incidence, respectively), and transmitted through the boundary applying the
appropriate Fresnel coefficients \cite{Jackson:1999aa}. Because we are dealing with a
point source and not plane waves, the standard transmission Fresnel coefficients must be
multiplied by a factor 
\be
\label{fresnelfactor}
 \frac{n_2}{n_1}\frac{\cos \theta}{\sqrt{1 - \left[\frac{n_2}{n_1}\sin \theta\right]^2}} ,
\ee
which can be understood as describing the change in divergence of rays upon transmission
\cite{James:2010vm}. The angle of refraction $\theta$ is the angle between the unit vector
$\hat{\boldsymbol{k}}$ and the normal to the boundary plane. The final transmission
formulas read
\ba
	\label{FresnelCoefficientsT}
	t_p &=& \frac{2 n_2 \cos \theta}{n_2 \sqrt{1 - \left[\frac{n_2}{n_1}\sin
	\theta\right]^2} + n_1 \cos \theta} ,\nonumber\\
	t_s &=& \frac{2 n_2 \cos \theta}{n_1 \sqrt{1 - \left[\frac{n_2}{n_1}\sin
	\theta\right]^2} + n_2 \cos \theta} .
\ea
When calculating the electric field at the observer position, {the $p$ polarization in the
reflection plane remains of course perpendicular to $\hat{\bf k}$, the new
propagation direction after refraction. }
In the following we refer to the electric field corresponding to tracks fully contained in
medium~1 as the ``transmitted contribution''.

For tracks fully contained in medium~2, two distinct contributions must be evaluated. The
first is the standard contribution given by Eq.~\eqref{ZHSFormula} describing the electric
field radiated into direction $\hat{\boldsymbol{k}}$ (``direct contribution''). The second
contribution (``reflected contribution'') comes from radiation emitted into a direction
$\hat{\boldsymbol{k}}'$ which is reflected off the boundary into the observing direction
$\hat{\boldsymbol{k}}$ (Fig.~\ref{fig:sketch_contributions}). Eq.~\eqref{ZHSFormula} is
also used to calculate this contribution. The electric field is again decomposed into $s$
and $p$ polarizations, and the appropriate Fresnel coefficients are used for reflection:
\ba
	\label{FresnelCoefficientsR}
	r_p &=& \frac{n_1 \cos \theta - n_2 \sqrt{1 - \left[\frac{n_2}{n_1}\sin
	\theta\right]^2}}{n_1 \cos \theta + n_2 \sqrt{1 - \left[\frac{n_2}{n_1}\sin
	\theta\right]^2}}, \nonumber\\
	r_s &=& \frac{n_2 \cos \theta - n_1 \sqrt{1 - \left[\frac{n_2}{n_1}\sin
	\theta\right]^2}}{n_2 \cos \theta + n_1 \sqrt{1 - \left[\frac{n_2}{n_1}\sin
	\theta\right]^2}}.
\ea
Since reflection does not change the divergence of rays, Eqs.~\eqref{FresnelCoefficientsR}
have their usual plane-wave form. As in the transmitted contribution, {the $p$ polarization
remains perpendicular to $\hat{\bf k}$ after reflection.}

Finally, for tracks crossing the two media the following procedure is applied.  The track
is split at the boundary plane into two sub-tracks, as in \cite{James:2010vm}. Each
sub-track is then fully contained in one medium and the methods previously described can
be applied to calculate the electric field. {For these tracks transition radiation is
naturally accounted for with this procedure.} In the ZHS algorithm each of the endpoints 
of a sub-track contributes a
term to the electric field.  For a charged particle moving uniformly in a homogeneous
medium ({\it i.e.} with constant velocity along the track),  the contributions from the
common endpoints of two adjacent sub-tracks cancel exactly. Thus, the electric field
associated to the track does not depend on the number or on the length of its sub-tracks.
On the other hand, a track crossing two media has two adjacent sub-tracks with common
endpoints located at the boundary. Since these endpoints are associated to sub-tracks
contained in media with different index of refraction, their contributions do not cancel
exactly.  Transition radiation appears because  of this incomplete cancellation. Notice
that the assumption of uniform velocity can be made to an arbitrary degree of precision by
reducing the length of the sub-tracks. 

The electric field produced by the entire shower is then obtained from the superposition
of the individual contributions of all particle tracks.  Explicit geometric phase
differences between tracks on both sides of the boundary plane are appropriately taken
into account in this procedure. This concludes the ZHS-TR algorithm.

The technical implementation of the ZHS-TR algorithm requires two simulation runs, since
the original ZHS code was developed for a single homogeneous medium. The simulation starts
by generating and propagating the shower in medium~1. Then, particles crossing the
boundary are propagated in medium~2 in a separate simulation. The ZHS code presents some
limitations, since it does not treat hadronic interactions and  radio emission induced by
the Earth magnetic field \cite{Huege:2003up}, and assumes a constant density of the media,
which is not appropriate for  atmospheric air. However, neglecting these effects should
not impact significantly our results as discussed in Sect.~\ref{sec:results}.

\subsubsection{Vacuum approximation}
\label{B2}

In the case under consideration, transition radiation is produced when the shower exits
into the atmosphere. Since the air density is much lower than the density of the medium
where the shower originated, medium~2 can be reasonably approximated by vacuum to simplify
calculations. There is no difference in the simulation for the portion of the shower
contained in medium~1: particle tracks are treated as in Sect.~\ref{B1}, and their
corresponding transmitted contribution is calculated. Instead, particles crossing the
boundary, which do not interact in vacuum, are modeled as moving to infinity with constant
velocity.  The electric field associated to such a semi-infinite track \cite{James:2010vm}
in the Fraunhofer regime is given by:
\be
	\label{EndpointsFormula}
	\boldsymbol{E}(\omega, \boldsymbol{x}) = - \frac{e \mu_r}{2 \pi \epsilon_0 c^2}i \omega \frac{e^{ikR}}{R}
	\frac{\boldsymbol{v}_\perp e^{i (\omega-\boldsymbol{k} \cdot \boldsymbol{v})
	t_1}}{i(\omega - \boldsymbol{k} \cdot \boldsymbol{v})} ,
\ee
where the notation follows Eq.~\eqref{ZHSFormula}. The origin of this formula can be
understood by dividing the semi-infinite track into sub-tracks. The electric field is
obtained by summing Eq.~\eqref{ZHSFormula}  over all sub-tracks. Since contributions of
the common endpoints of adjacent sub-tracks cancel, the only term left corresponds to the
starting point (the term at infinity goes as $1/R \rightarrow 0$), thus giving
Eq.~\eqref{EndpointsFormula}. 

Notice that since the electric field of the entire track is given by
Eq.~\eqref{EndpointsFormula}, there is no need for a division in sub-tracks, which makes
the simulation faster. Equation~\eqref{EndpointsFormula} is then used to calculate the
direct and reflected contributions to the electric field following the procedure of
Sect.~\ref{B1}.  We will apply the vacuum approximation in Sect.~\ref{vertical2}.

\subsection{Transition radiation from a single particle}
\label{sec:Single}

\begin{figure}
\center
\includegraphics[width = 0.48 \textwidth]{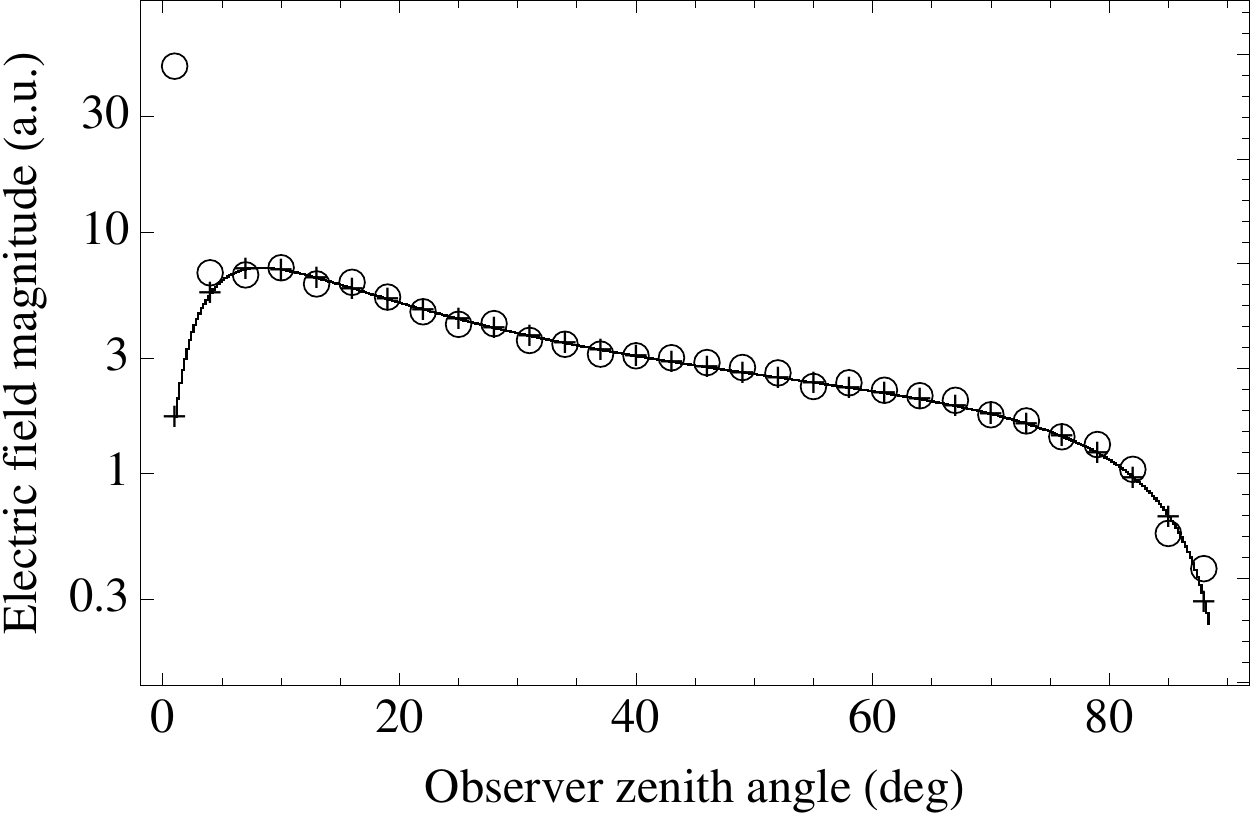}
\caption{
Angular distribution of the far-field electric field radiated by a single charged particle
crossing from ice to air: comparison of the results obtained using the ZHS-TR 
algorithm with sub-tracks 10~cm (crosses) and 100~cm (open dots) long to the exact 
calculation (black line).
}
\label{fig:single_particle}
\end{figure}

A basic check of the ZHS-TR algorithm was performed by estimating the transition radiation
from a single charged particle and comparing the result with the classical derivation
\cite{Ginzburg:1990aa}. An electron crossing the boundary from ice to air in the vertical
direction was considered for this purpose. The electron velocity was fixed to $0.99c$,
corresponding to an energy of about 3.6 MeV. In order to approximate an infinite track
(required for the classical derivation), electron interactions were not considered, and a
total track length of 100~km was equally divided between the two media. We then evaluated
the electric field at an observer position $\bf d$ of zenith angle $\theta$, and $|{\bf
d}|=1$~km (the origin of the coordinate system coincides with the intersection of the
track with the boundary plane). Since this distance is much larger than the length of the
sub-tracks in ZHS-TR, we expect the Fraunhofer limit used in Eq.~\eqref{ZHSFormula} to be
a good approximation. 

In Fig.~\ref{fig:single_particle}, the electric field calculated with our algorithm is
compared with the exact result obtained by solving the Maxwell equations for an infinite
track \cite{Ginzburg:1990aa}. The two approaches are in practically perfect agreement when
sub-tracks of 10~cm length are used in ZHS-TR. To illustrate the importance of the
Fraunhofer limit, we also show the same comparison for sub-tracks of 100~cm length. At
small zenith angles, where a discrepancy is observed, the perpendicular distance between
the track and the observer ($\sim |{\bf d}| \theta$) becomes comparable to the length of
the sub-tracks, and the Fraunhofer regime is no longer valid. 

In the rest of the paper, we will assume that the observer is always distant enough from
the shower for the Fraunhofer regime to be valid. Due to the finite extent of the shower,
this is a good approximation in most practical cases, for example in the detection of the
TR signal from a satellite or balloon instrument \cite{Gorham:2008dv},
\cite{Motloch:2013kva}.

\section{Properties of transition radiation from high-energy showers}
\label{sec:results}

In this section, we investigate in detail the properties of the radiation emitted by an
electromagnetic shower transitioning from a dense medium to air.  Showers initiated by a
100~TeV electron in ice ($n_1=1.78$ at radio frequencies) were used in most of these
studies. Also, no thinning was employed in the simulation of the showers ({\it i.e.} all
particles were propagated). Since in the Fraunhofer
approximation the electric field ${\bold E}(\omega,{\bf x})$ is inversely proportional to
the distance $R$ to the shower, results are presented in terms of the quantity $|R \times
{\bold E}|$. The case of a vertical shower is treated first, and then results for inclined
showers - which may be more realistic for a UHE neutrino-induced interaction - are
presented.     
 
\subsection{Vertical showers} \label{subsec:vertical}

\subsubsection{Direct, reflected and transmitted contributions}
\label{vertical1}
\begin{figure*}
\center
\includegraphics[width = \textwidth]{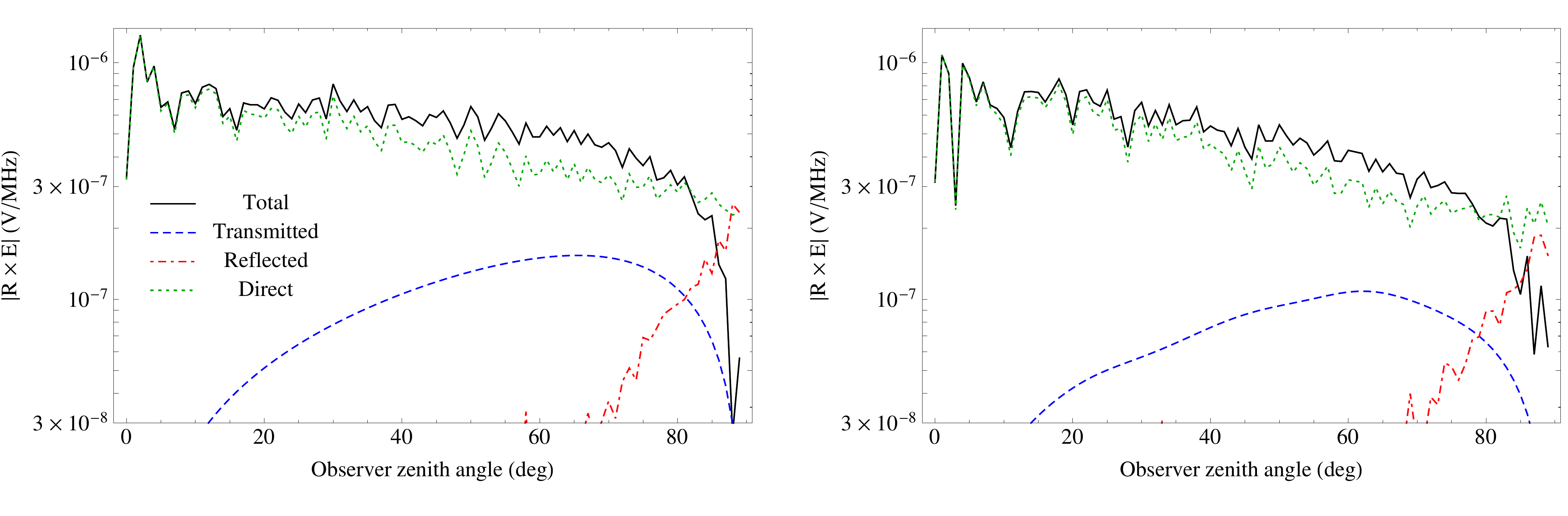}
\caption{
 Angular distribution of the electric field radiated by a single 100 TeV vertical
 shower crossing from ice to air: scaled magnitudes of the total field (black), direct
 (green short dashes), reflected (red dot-dashes) and transmitted (blue dashes)
 contributions. Due to phase differences the total magnitude can be smaller than those of
 individual parts. See the text for details. The electric field is evaluated at 200~MHz
 (left panel) and 1~GHz (right panel). 
 } 
 \label{fig:comparison_DRT} 
 \end{figure*}

The angular distribution of the electric field radiated by a 100~TeV vertical shower
simulated with ZHS-TR is shown in Fig.~\ref{fig:comparison_DRT} (black solid line). The
shower's starting point in ice was chosen appropriately for the shower maximum to occur
approximately at the ice--air boundary. The zenith angle
dependence reflects the contributions to TR by single electrons whose angular features are
shown in Fig.~\ref{fig:single_particle}. The weak angular dependence of the emitted
radiation is a notable feature, when compared to coherent Cherenkov radiation which is
strongly beamed around the Cherenkov angle \cite{Zas:1991jv}. 
A value of $|R \times {\bold
E}| \sim 5 \times 10^{-7}\, \mathrm{V/MHz}$ is obtained for this shower.

\begin{figure*}
\center
\includegraphics[width = \textwidth]{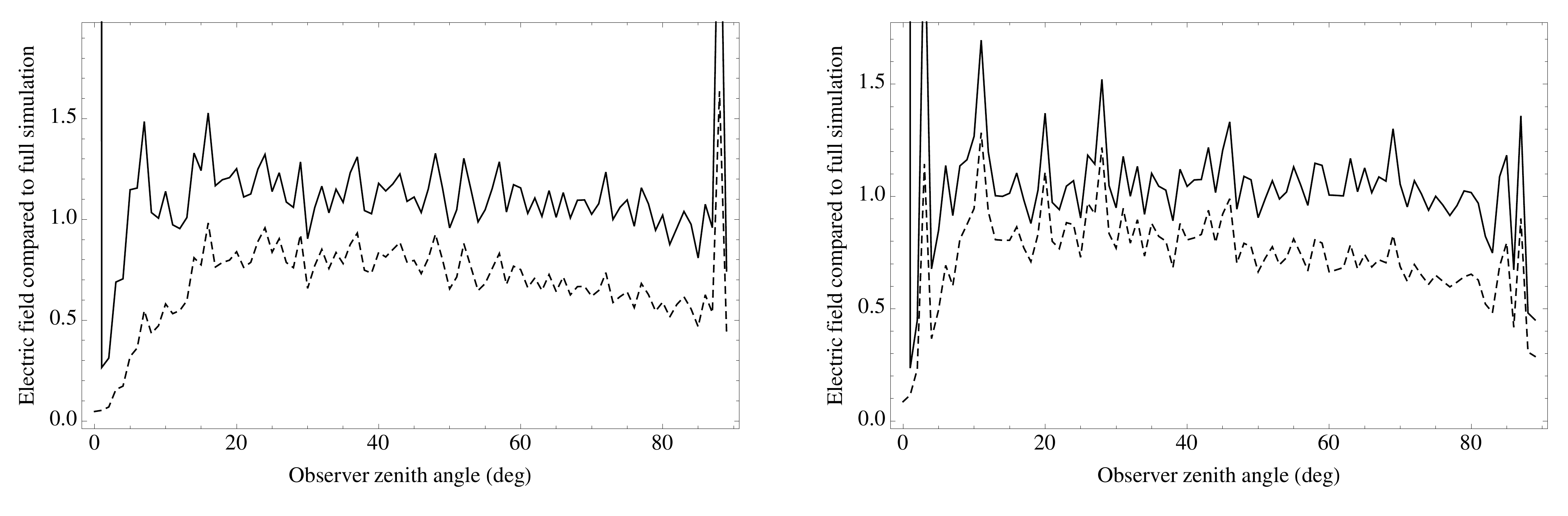}
\caption{
Ratio of the electric field magnitudes calculated in the vacuum approximation and 
the full simulation (solid) and the same comparison for simulation where we include only tracks
crossing the ice--air boundary (dashed). 
The vertical shower started with 100 TeV energy and crossed from ice to air.
The electric field is evaluated at 200~MHz (left panel) and 1~GHz (right panel).
}
\label{fig:approximations}
\end{figure*}

Also shown in Fig.~\ref{fig:comparison_DRT} are the contributions to the total electric
field (direct, transmitted and reflected) introduced in Sect.~\ref{B1}. The direct
contribution from particle tracks in air is dominant at all angles, and presents a peak at
$\sim 2^\circ$. This peak can be understood as Cherenkov radiation produced in air
(Cherenkov angle $\sim1.4^\circ$) by the high energy shower particles which follow closely
the shower axis. In addition, there are noticeable fluctuations around the mean value of
$|R \times {\bf E}|$, which are due to the fine structure of the shower development in
air. In fact, the spatial dimension corresponding to the frequencies of observation -
200~MHz and 1~GHz - is much smaller than the transverse size of the shower (a simulation
performed at 10~MHz, not shown for the sake of brevity, presents a much smoother
behavior).  The reflected contribution is negligible except for the largest angles, where
it becomes comparable to the direct contribution.  However, the two contributions cancel
each other for an observer looking orthogonally to the shower axis, since a $\pi$ phase
shift is gained upon reflection, and the total electric field drops at zenith angles
approaching $90^\circ$. Lastly, the transmitted contribution from particles in ice is
never dominant for a vertical shower. In fact, the coherent Cherenkov radiation from the
Askaryan effect in ice undergoes total internal reflection at the ice--air boundary plane
(the Cherenkov angle is larger that the critical angle for $n_1$=1.78) and cannot reach
the observer. However, this component can become significant for inclined showers, see
Sect.~\ref{subsec:inclined}. 

\subsubsection{Particles crossing the boundary and vacuum approximation}
\label{vertical2}
To further investigate the origin of the emitted radiation, we performed a dedicated
ZHS-TR simulation where the electric field was calculated taking into account only
contributions from particles crossing the ice -- air boundary. Results are shown in
Fig.~\ref{fig:approximations}, where we plot the ratio of the electric field magnitudes
obtained in this simplied simulation and the electric field magnitudes of the 
full result. Indeed, transition radiation from particles crossing the
boundary accounts for most of the emission (at low zenith angles, Cherenkov radiation from
particles propagating in air is dominant, as seen in Fig.~\ref{fig:comparison_DRT}).  

Also shown in Fig.~\ref{fig:approximations} is a ratio of the electric field
magnitudes resulting from a ZHS-TR simulation using the
vacuum approximation of Sect.~\ref{B2} with the electric field magnitudes of the full
result. The vacuum approximation reproduces the full
ZHS-TR simulation reasonably well, with a notable exception at low zenith angles. This can be
explained by the fact that, for an observer looking from the vertical direction, highly
relativistic particles with $\boldsymbol{k} \parallel \boldsymbol{v}$ have $\omega -
\boldsymbol{v}\cdot\boldsymbol{k} \sim 0$, resulting in a large contribution from
Eq.~\eqref{EndpointsFormula}. High energy particles emerge from ice strongly beamed along
the vertical axis, and since their direction does not change in the vacuum approximation,
a strong peak is expected. On the other hand, the full ZHS-TR simulation includes particle
interactions in air which randomize the directions of the sub-tracks, leading to a
suppression of the peak. In any case, the agreement between the vacuum approximation
and the full simulation over most of the zenith angles suggests that the details of the
shower development in air have a marginal effect on the TR emission. Thus, we expect that
neglecting the altitude dependence of the air density and geomagnetic effects in particle
propagation will not change significantly our estimates of transition radiation. 

\subsubsection{Frequency spectrum}
\label{vertical4}
\begin{figure}
\center
\includegraphics[width = 0.48 \textwidth]{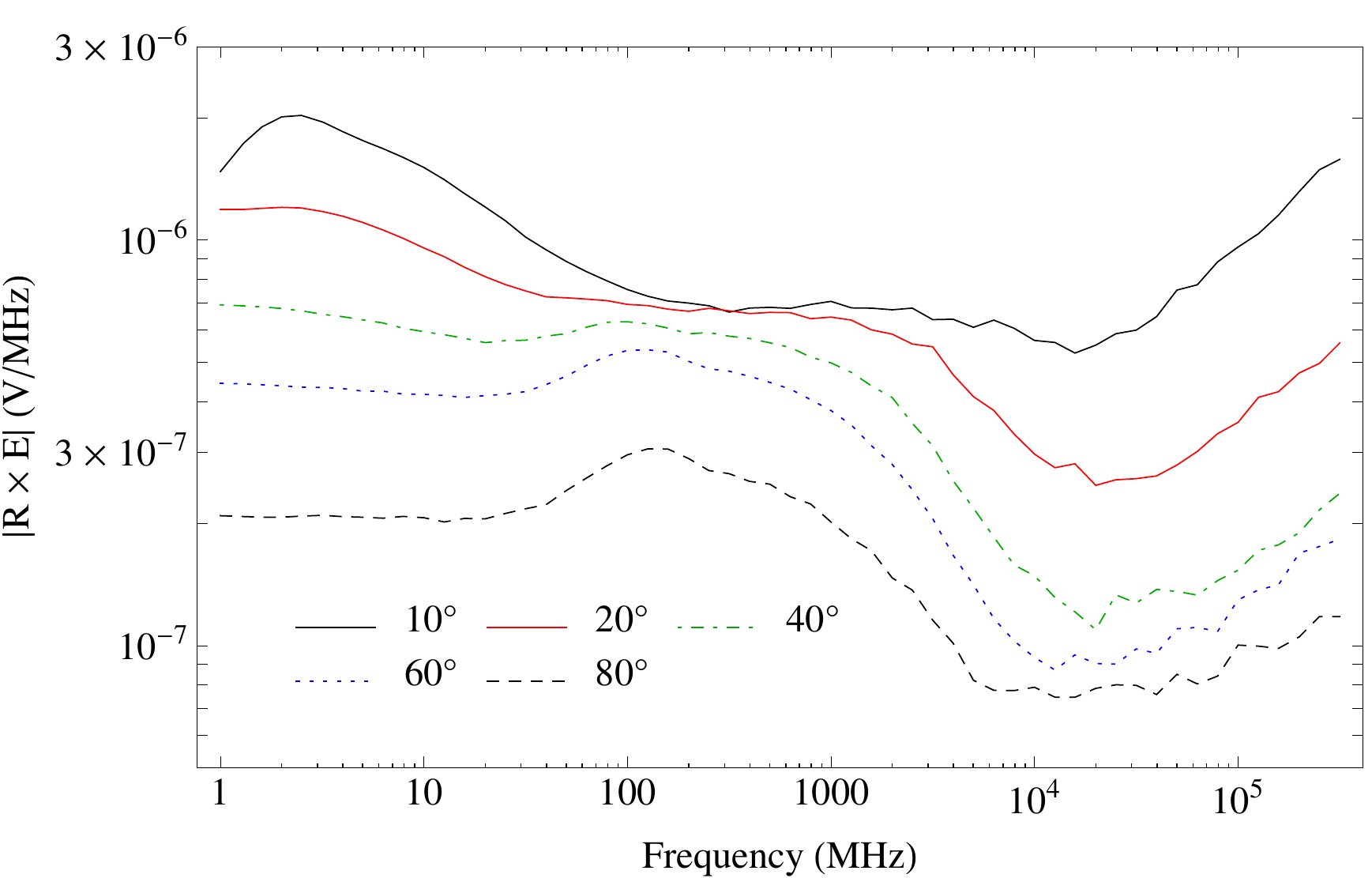}
\caption{Frequency spectrum of the electric field radiated by a 100 TeV vertical shower
crossing from ice to air, at $10^\circ$ (black solid), $20^\circ$ (red solid), $40^\circ$
(green dot dashes), $60^\circ$ (blue short dashes), and $80^\circ$
(black long dashes) zenith angle of observation (top to bottom).
Each curve is an average of a hundred showers. The rise at the highest frequencies
comes from incoherent emission.}
\label{fig:frequency_dept}
\end{figure}

Another important property of the radiation is its frequency spectrum, shown in
Fig.~\ref{fig:frequency_dept} for 100~TeV showers. To reduce shower-to-shower
fluctuations, a hundred showers were simulated, and the modulus of their signal was
averaged. All averages will be done this way unless explicitly stated. The rise at
the highest frequencies is caused by contributions from incoherent emission and continues
well beyond the frequencies displayed. To check that the coherent contribution is indeed
quickly dying away after certain threshold frequency, we fitted a power law dependence to
the magnitude of the electric field above 20~GHz where the coherent contributions are
negligible. When we subtracted this contribution from the total electric field observed in the
simulations, the remainder showed the expected steep drop off.

Unlike  coherent
Cherenkov radiation, which drops off at relatively low frequencies away from the Cherenkov
angle \cite{Zas:1991jv},  transition radiation is coherent up to $\sim 1\, \mathrm{GHz}$
over a wide range of angles.  The physical origin of this effect may be understood with a
simplified model where TR is produced only by particles crossing orthogonally the boundary
plane with velocity $c$, uniformly distributed over a disk of radius $r_M \sim$ 11~cm (the
Moli\`{e}re radius in ice).  Under these approximations, each particle contributes equally
in amplitude, and only the relative phases due to propagation are important. The problem
then becomes analogous to calculating the diffraction pattern of a uniformly-illuminated
circular aperture \cite{Jackson:1999aa}. Thus, the electric field at given frequency $f$ goes as
\be
	|{\bf E}| \sim \frac{c}{f \sin\theta}J_1\(\frac{2 \pi f r_M}{c} \sin \theta\),
\ee
where $J_1$ is a Bessel function of the first kind.
This expression is maximal for $f \rightarrow 0$ and reaches the first zero when 
\be
\label{eq:cutoff}
	f \sim \frac{0.61 c}{r_M\sin \theta}.
\ee
This equation may be taken as an estimate of the TR cutoff frequency. Indeed, the trend
observed in Fig.~\ref{fig:frequency_dept}, with lower cutoff frequencies corresponding to
larger angles of observation, is well described qualitatively by Eq.~\eqref{eq:cutoff}.

\subsubsection{Scaling with number of particles crossing the boundary}

We have shown in Sect.~\ref{vertical2} that most of the emission from a shower emerging
from ice into air is due to particles crossing the boundary. These particles should
contribute coherently to the radio signal up to a cutoff frequency related to the
geometrical size of the shower. Thus, we expect the averaged electric field $\langle R
\times E \rangle$ to scale linearly with the charge excess $\Delta Q$ (the number of
electrons minus the number of positrons) flowing through the boundary. 

To test this we used an averaged value 
\ba
	\langle E \times R \rangle &=& \frac{1}{C}\int_{10^\circ < \theta < 70^\circ} 
	\left| R \times E(200\, \mathrm{MHz},\theta)\right|\mathrm{d}\Omega \nonumber\\
	&=& \frac{1}{C'} \int_{10^\circ}^{70^\circ}
	\left| R \times E(200\, \mathrm{MHz},\theta)\right|\sin \theta\mathrm{d}\theta ,
\ea
where the normalization constants are
\be
	C' = \frac{C}{2\pi} = \int_{10^\circ}^{70^\circ} \sin \theta\mathrm{d}\theta .
\ee
This quantity measures an average magnitude of the electric field in the middle angular
region which is not affected by the air Cherenkov emission or the decrease of the signal
at large zenith angles.

\begin{figure}
\center
\includegraphics[width = 0.48 \textwidth]{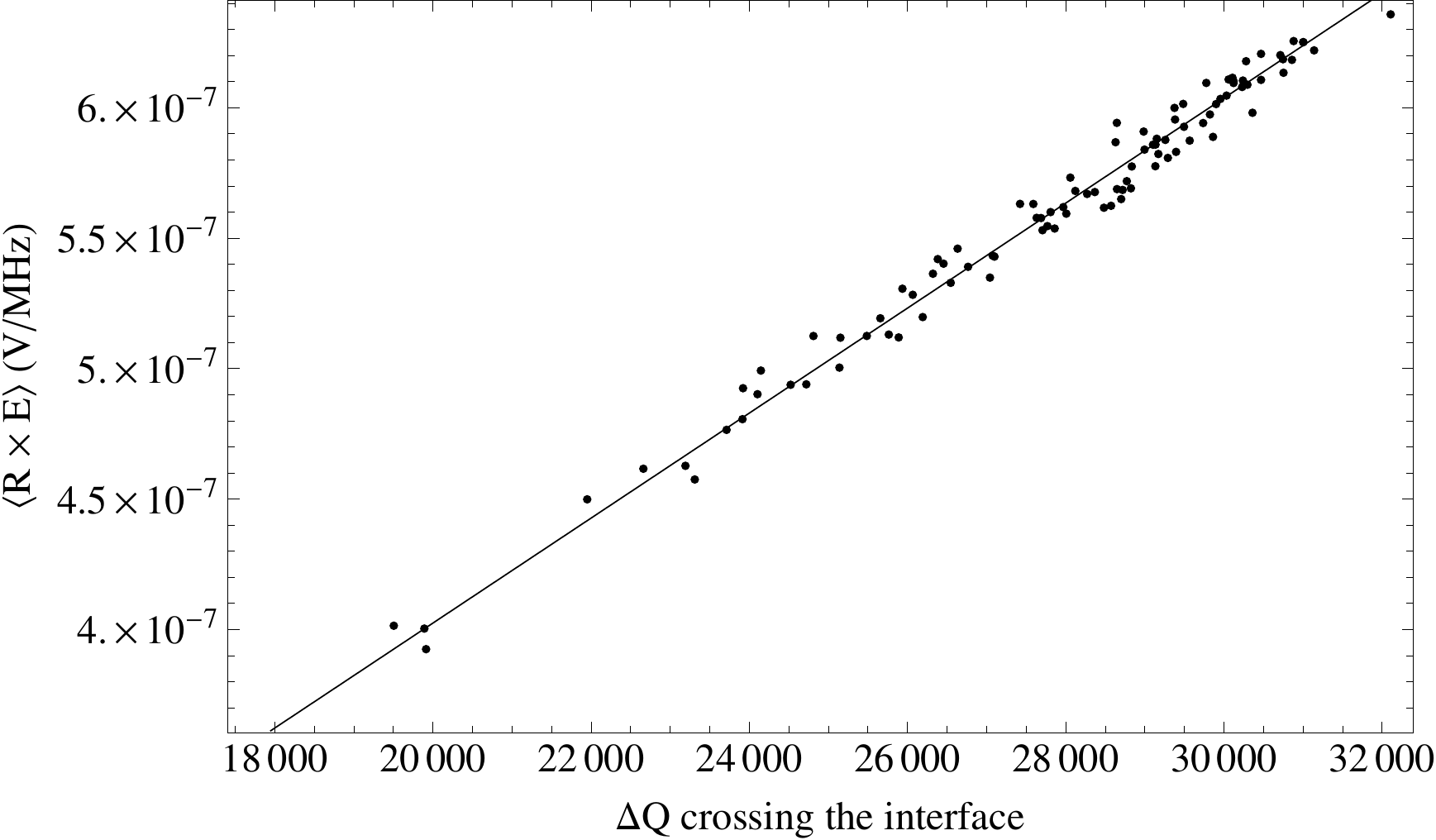}
\caption{Relation between the absolute charge excess crossing the boundary and the
averaged electric field for a hundred 100~TeV showers starting 500 g/cm${}^2$ from the ice-air
boundary. The straight line is a fit of the expected theoretical dependence.}
\label{fig:nCross}
\end{figure}

The theoretically predicted linear scaling is confirmed in Fig.~\ref{fig:nCross}, which
shows the dependence of the average magnitude of the electric field as a function of the
charge excess for a hundred simulated 100 TeV showers starting at a fixed distance $500\,
\mathrm{g/cm^2}$ from the ice-air boundary. One can notice the presence of showers with
charge excess of only 60\% of the maximum observed value. This effect is related to the
fact that while for most of the showers $500\, \mathrm{g/cm^2}$ corresponds approximately
to the distance to the shower maximum, for others the shower maximum occurs at a somewhat
different distance.  This affects the number of crossing particles and relatedly the size
of the mean electric field.  

On top of the systematic increase and decrease of the mean value of the electric
field with the charge excess through the boundary there is also the effect of the fine
structure of an individual shower. This leads to fluctuations around the mean value, seen
for example in Fig.~\ref{fig:comparison_DRT}. From this figure we can estimate the magnitude
of these fluctuations for a 100 TeV shower; for showers of higher energies these
fluctuations are relatively smaller due to higher total number of particles in these
showers.

\begin{figure*}
\center
\includegraphics[width = \textwidth]{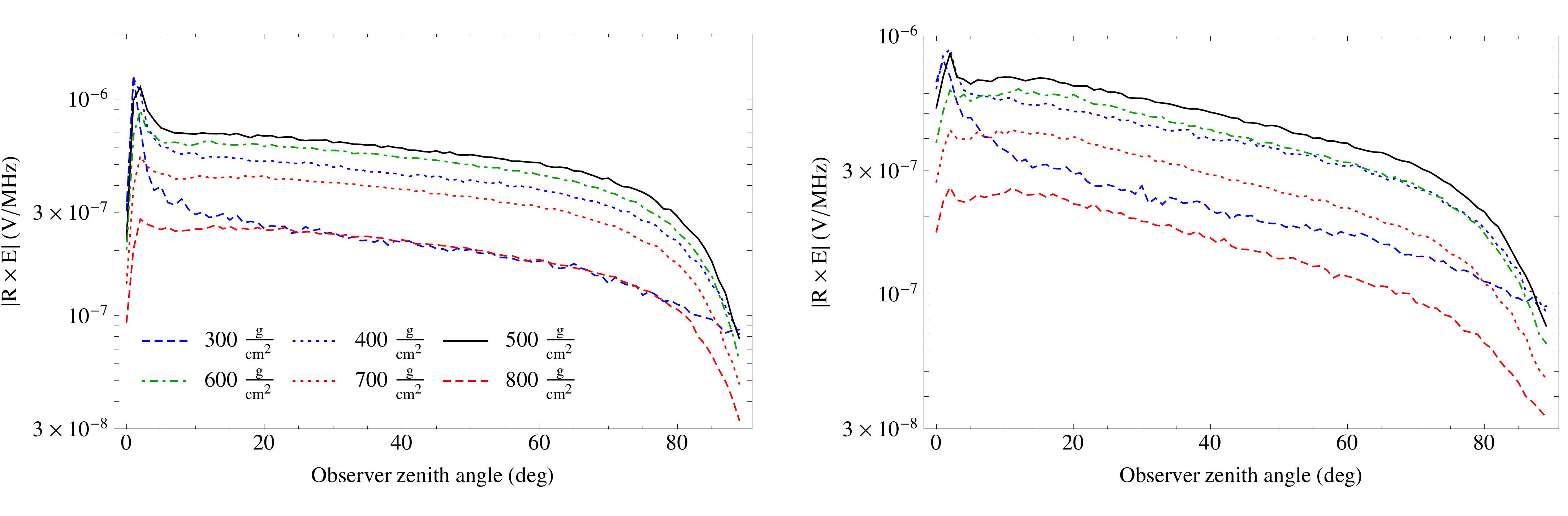}
\caption{Angular distribution of the electric field radiated by a 100~TeV vertical shower
crossing from ice to air, with starting point in ice at 300 (blue long dashes), 400 (blue
short dashes), 500
(black, showers typically crossing at shower maximum), 
600 (green dot-dashes), 700 (red short dashes) and 800 (red long dashes) $\mathrm{g/cm^2}$ from the ice--air boundary.
The electric field is evaluated at 200~MHz (left panel) and 1~GHz (right panel). Each
curve is an average of a hundred showers.}
\label{fig:depth}
\end{figure*}

In another related study, we simulated 100~TeV showers starting at different ice depths.
Since the number of particles crossing the boundary depends in this case on the stage of
the shower development, the signal should be largest when the maximum of the shower
development occurs close to the boundary for a typical shower. 
This is indeed confirmed in Fig.~\ref{fig:depth}. The
peak at low observation angles ($< 5^\circ$), which is due to Cherenkov radiation from
particles in air, becomes less pronounced the farther the shower starting point is from
the boundary. In fact, at later stages of the shower development the shower is less
compact and its particles have a larger angular spread, resulting in a loss of coherence.

\subsubsection{Dependence on shower particles energy spectrum}
\label{vertical5}
\begin{figure*}
\center
\includegraphics[width = \textwidth]{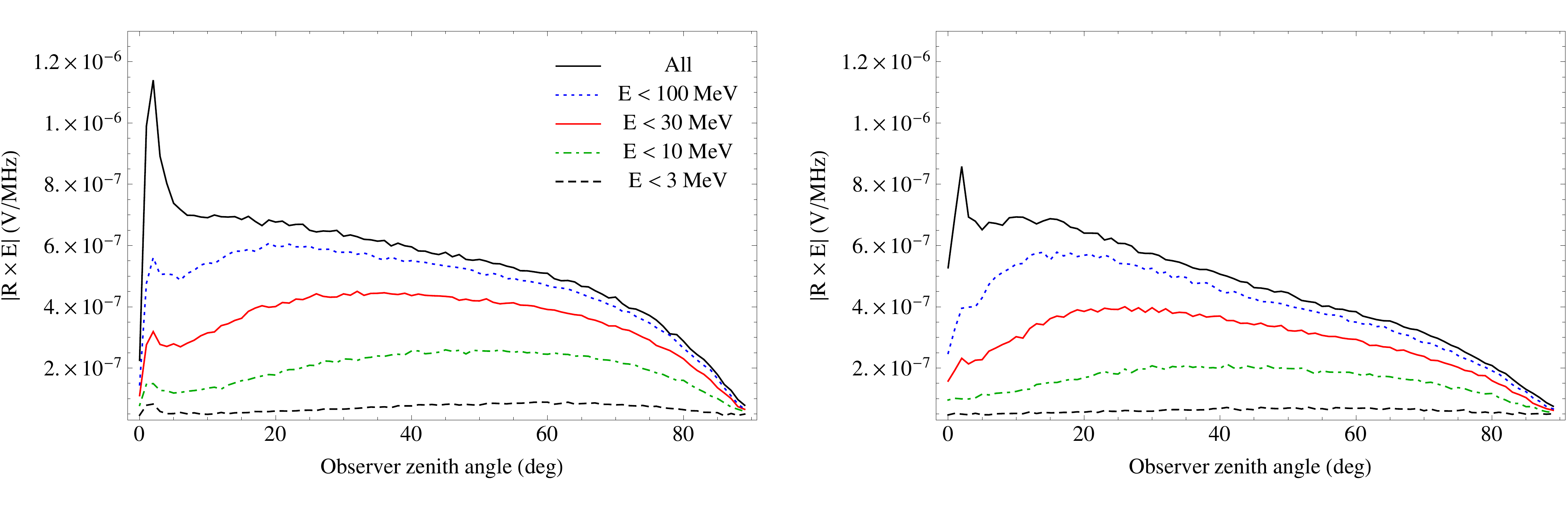}
\caption{Angular distribution of the electric field radiated by a 100 TeV vertical shower
crossing from ice to air (from top to bottom): all particles (black), particles with energy $<$100 MeV (blue
short dashes),
30 MeV (red), 10 MeV (green dot dashes) and 3 MeV (black long dashes). The electric field is evaluated at
200~MHz (left panel) and 1~GHz (right panel). Each curve is an average of a hundred showers.}
\label{fig:energy_decomposition}
\end{figure*}

Since the electric field depends on the particle velocity ({\it c.f.}
Eq.~\eqref{ZHSFormula}), we may expect a dependence of the TR  on the energy spectrum of
the shower particles. To study this dependence, we estimated the electric field for
100~TeV simulated showers taking into account only the contributions of sub-tracks from
particles with energy below a given threshold. Results are presented in
Fig.~\ref{fig:energy_decomposition} (note the linear scale of the vertical axis). The
highest energy particles ($> 100$~MeV) contribute mostly at low zenith angles. In fact,
these particles are strongly beamed in the vertical direction, and account for the
coherent Cherenkov peak as discussed in Sect.~\ref{vertical1}. The contributions from
particles in the energy ranges 3--10 MeV, 10--30 MeV, and  30--100 MeV are comparable.
This is due to a compensation between the number of particles  -- which is higher at low
energies -- and the amount of radiation coming from a single particle -- which is higher
at high energies. Particles with energies below 3~MeV do not contribute significantly.

\subsubsection{Dependence on the shower energy}

We expect the absolute charge excess for showers which cross the boundary at their
maximum developments to be a good proxy of the shower energy as the number of particles at
the shower maximum is roughly proportional to the shower energy. As a consequence, we
expect the electric field to scale linearly with the shower energy. 

In our implementation of the ZHS-TR algorithm it is not possible to know a priori whether a given
simulated shower will hit the boundary around the shower maximum or whether it will be one
of the downward fluctuations seen in the Fig.~\ref{fig:nCross}. To quantitatively
asses how the radiated electric field depends on the shower energy, we then compare
maximal signal seen in all our simulated showers at three different energies: 1, 10 and 100
TeV. As a first step we run several trial showers to find for each of these energies an
approximate starting point where we can expect showers reaching the boundary
at roughly their shower maximum. We then simulate a hundred unthinned
showers starting at this point and search for the maximal value of $\langle E \times R
\rangle$. We take this quantity (denoted as $\langle E \times R \rangle^{\rm max}$ as our
proxy for the value of $\langle E \times R\rangle$ for a shower hitting the boundary at
the shower maximum. Similar other choices are possible but lead all to comparable results.

\begin{figure*}
\center
\includegraphics[width = \textwidth]{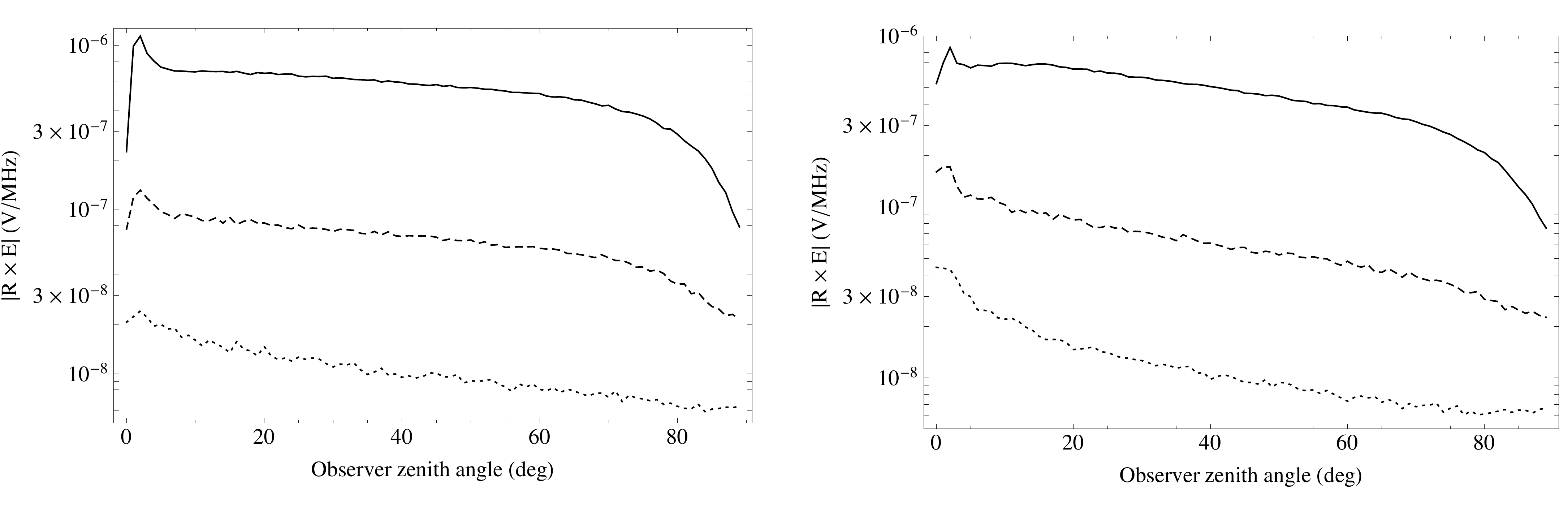}
\caption{Angular distribution of the electric field radiated by 1 (short dash), 10 (long
dash) and 100~TeV (solid) vertical shower crossing from ice to air. The electric field is
evaluated at 200~MHz (left panel) and 1~GHz (right panel). Each curve is an average of
a hundred showers.}
\label{fig:energy}
\end{figure*}

The maximal values of the mean electric field at these three energies observed in our
sample of showers are
\ba
	\langle E \times R \rangle^{\rm max}_{100\,\mathrm{TeV}} &=& 6.4 \times 10^{-7}\, \mathrm{V/MHz}\\\nonumber
	\langle E \times R \rangle^{\rm max}_{10\,\mathrm{TeV}} &=& 7.8 \times 10^{-8}\,
	\mathrm{V/MHz}\\\nonumber
	\langle E \times R \rangle^{\rm max}_{1\,\mathrm{TeV}} &=& 1.2 \times 10^{-8}\,
	\mathrm{V/MHz}.
\ea
They deviate slightly from the theoretical prediction, which we attribute to higher effect
of fluctuations at lower energies. This viewpoint is supported by Fig. \ref{fig:energy},
which shows the angular distributions of the electric fields averaged over the investigated
showers. While the 100~TeV showers show profiles comparable to those of a single particle,
at lower energies the fluctuations lead to a change in the shape of the profile. Notice
that even for 100~TeV showers the agreement with the single particle profile is worse at
higher frequencies due to the lower level of coherence.

As a further study, we investigated several thinned 1~EeV showers. After finding an
approximate shower development $D$ to the shower maximum we then repeated the
simulation for showers with starting point separated by distance $D$ from the
interface. Due to increased computational demands we simulated only twelve showers for
this energy. The largest value of $\langle E \times R \rangle$ we found among these showers is 
\be
	\langle E \times R \rangle^{\rm max}_{1\,\mathrm{EeV}} = 1.52\,\mathrm{mV/MHz} \approx 2400\,
	\langle E \times R \rangle^{\rm max}_{100\,\mathrm{TeV}} .
\ee 
The reason why the electric field at very high energies does not scale linearly with the
energy can be traced back to the Landau-Pomeranchuk-Migdal (LPM) effect
\cite{Landau:1953um, Landau:1953gr}. As was found in \cite{AlvarezMuniz:1997sh}, due to
elongation of the shower at EeV and higher energies the number of particles around shower
maximum for a shower with the LPM effect included decreases with respect to the same
shower with no LPM effect. Consequently, the charge excess and the radiated electric field
are also lowered.

\subsubsection{Different media}
\label{vertical6}

\begin{figure*}
\center
\includegraphics[width = \textwidth]{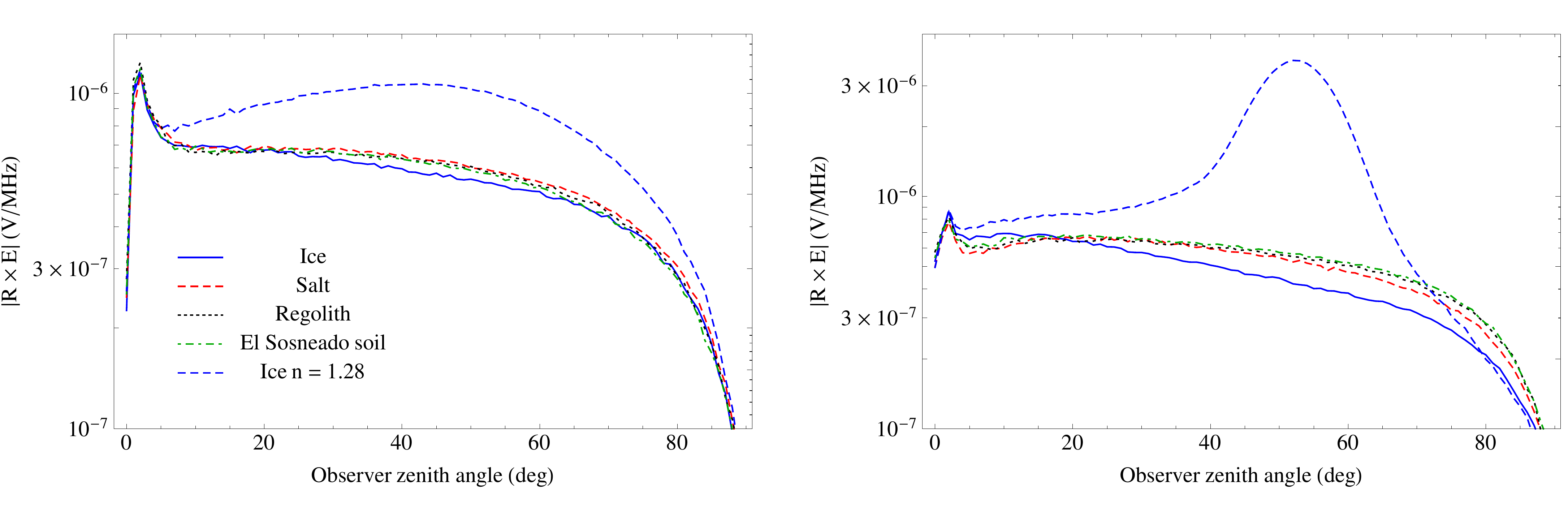}
\caption{Angular distribution of the electric field radiated by a 100~TeV vertical shower
crossing from ice (blue), salt (red dashes), regolith (black short dashes) and soil at El Sosneado
(green dot-dashes) to air. 
Also shown is the electric field for a simulation where the ice refractive index was set
to the unphysical value of 1.28 (blue long dashes). 
The electric field is evaluated at 200~MHz (left panel) and 1~GHz (right panel). Each
curve is an average of a hundred showers.}
\label{fig:material}
\end{figure*}

So far, we have studied the properties of TR from showers crossing an ice--air interface.
Salt, soil and regolith are also potentially relevant for  an experiment, and we performed
simulations of 100~TeV showers starting in these media. For soil we took parameters of the soil from 
the Argentinian village El Sosneado, next to the Pierre Auger Observatory location. The
angular distribution of the electric field radiated in salt/regolith/soil simulations is
compared with the standard ice simulation in Fig.~\ref{fig:material}. At 200~MHz, the
electric field is comparable for all four media. At 1~GHz, a lower emission is obtained in
ice.  This can be explained by the larger Moli\`{e}re radius of this medium ($\sim$
roughly double of the other $r_M$), which results in a less compact shower. Since the
condition of coherent emission from the shower particles is then more easily broken, a
weaker electric field is expected. Soil and lunar regolith have rather similar behaviors
at both frequencies. 

To illustrate an additional feature which can occur for certain configurations, 
we performed a simulation, also shown
in Fig.~\ref{fig:material}, in which the refractive index of a medium, otherwise similar
to ice, was artificially set to 
value of 1.28 so that the Cherenkov angle is smaller than the critical angle for total internal
reflection in this medium (see Sect.\ref{vertical1}) and the coherent Askaryan radiation
produced by the shower in the artificial ice is now transmitted into air (see
Sect.\ref{vertical1}).  Notice that the refracted Cherenkov radiation is significantly
larger than the TR, and presents a characteristic angular distribution, with a beam
clearly visible at 1~GHz where coherence effects are stronger.  A signal of similar origin
will be encountered in the study of inclined showers in Sect.~\ref{subsec:inclined}.
 
\subsection{Inclined showers}
\label{subsec:inclined}
While some specific applications of transition radiation from close-to-vertical showers
may be found (for example, showers emerging from a mountain range), a more likely use of
TR could be the detection of UHE neutrinos through inclined showers emerging from the
Earth surface. In fact, the Earth becomes increasingly opaque for neutrino energies above
100~TeV \cite{Gandhi:1998ri}, and most of the observable showers would originate from
Earth-skimming neutrinos interacting below the Earth surface.   

\begin{figure*} \centering \begin{minipage}[t]{.49\textwidth}
\includegraphics[width=\textwidth]{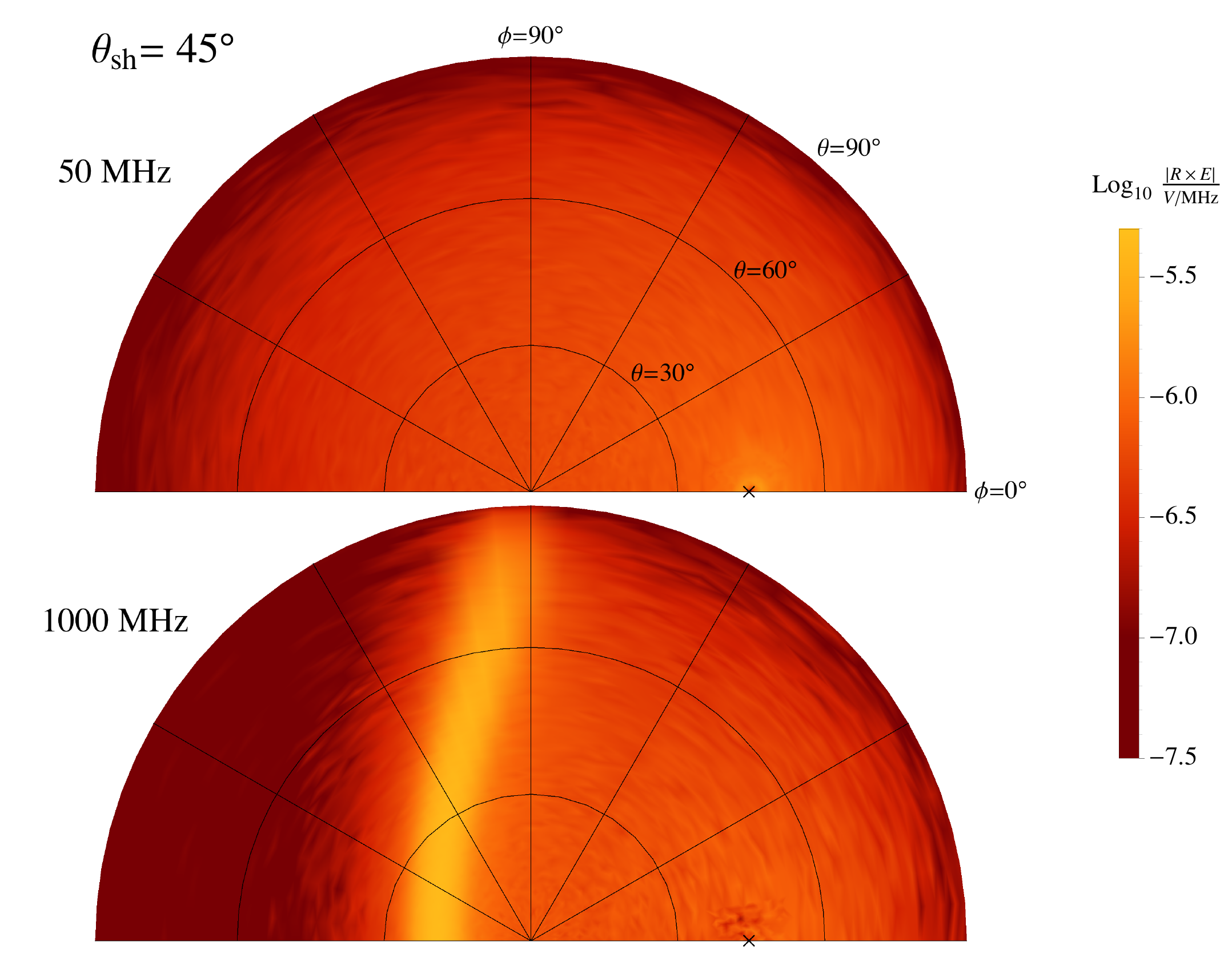} 

\caption{Angular distribution of the electric field radiated by a 100~TeV inclined shower
crossing from ice to air with incident zenith angle of $\theta_{\rm sh} = 45^\circ$ and
$\phi_{\rm sh} = 0^\circ$. Each point corresponds
to the electric field in a direction $(\theta,\phi)$ of observation, 
the polar angle is equal to $\phi$ and the radial distance from the origin
proportional to $\theta$. The shower leaves medium 1 (ice) into medium 2 (air) while crossing
at shower maximum. Due to symmetry of the situation we show ony half of the sky in
each plot. The direction of the shower axis is represented by a cross.  The
electric field is evaluated at a frequency of 50~MHz (top) and 1~GHz (bottom).} 
\label{fig:inclined45} 
\end{minipage}
\hfill 
\begin{minipage}[t]{.49\textwidth}
\includegraphics[width = \textwidth]{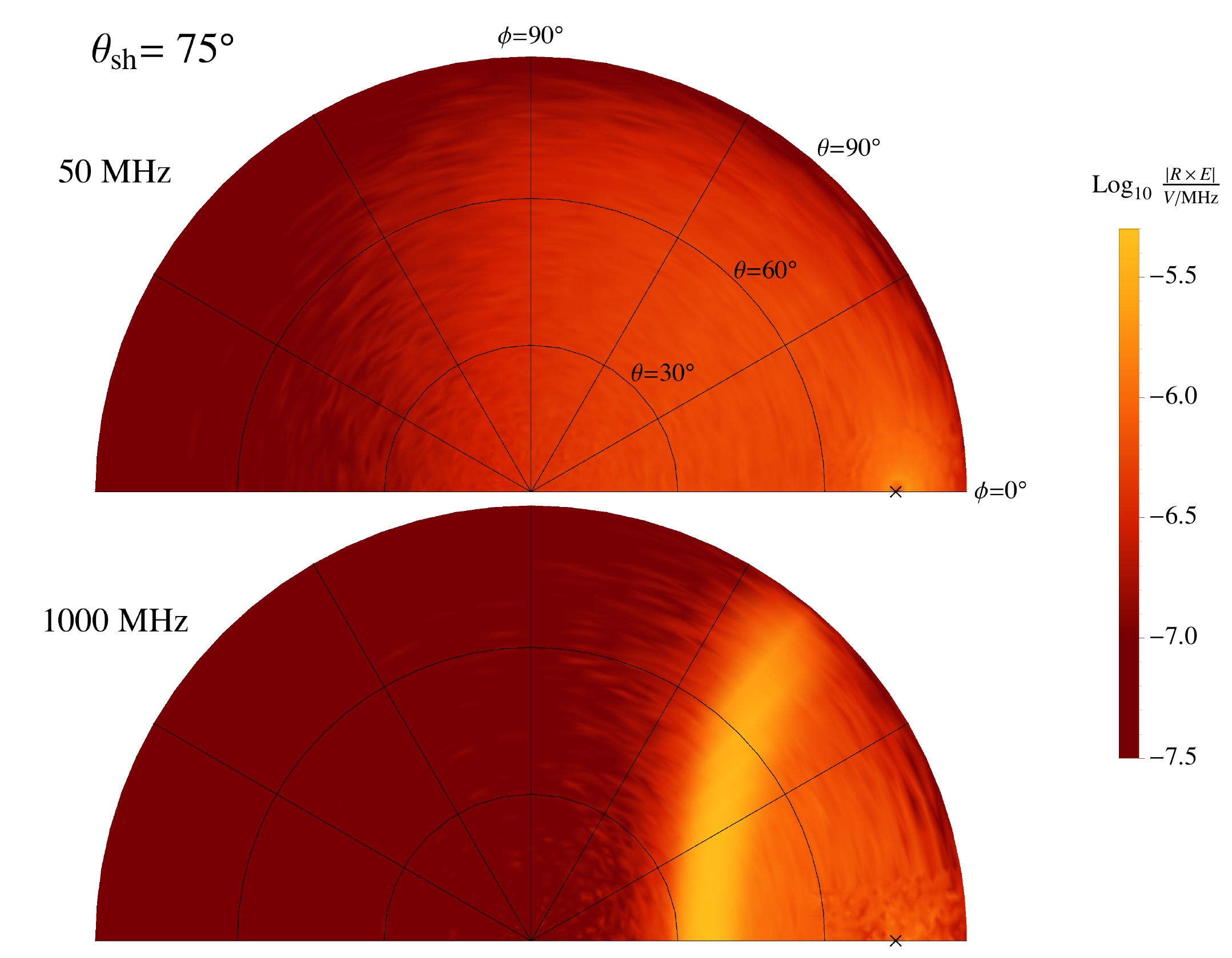}
\caption{{Same as Fig.~\ref{fig:inclined45} for a shower with incident zenith angle of
$\theta_{\rm sh} = 75^\circ$ and $\phi_{\rm sh} = 0^\circ$.}}
\label{fig:inclined75}
\vfill
\end{minipage}
\end{figure*}

\begin{figure*} \centering \begin{minipage}[t]{.49\textwidth}
\includegraphics[width=\textwidth]{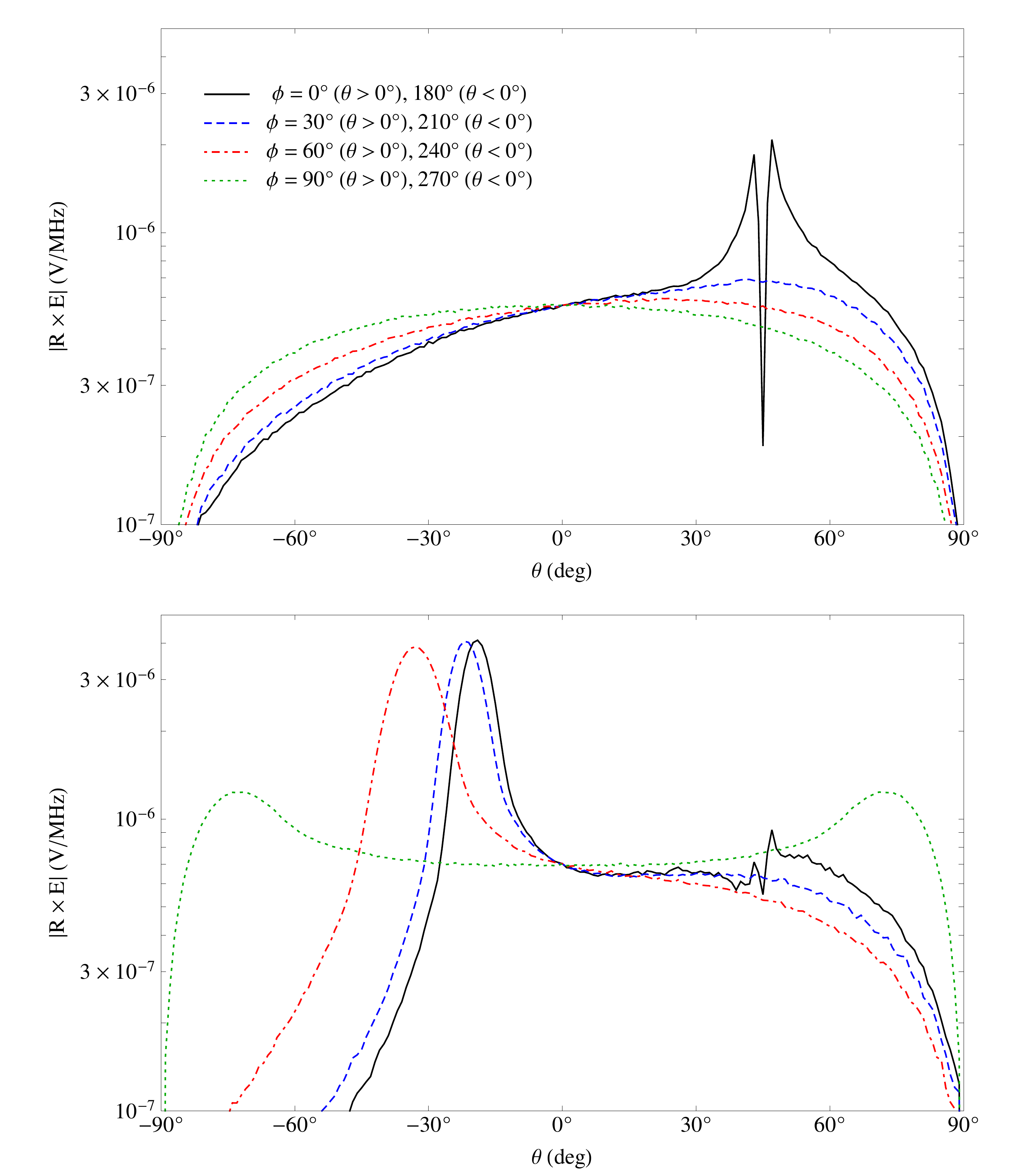} 

\caption{Angular distribution of the electric field radiated by a 100~TeV inclined shower
crossing from ice to air with incident zenith angle of $\theta_{\rm sh} = 45^\circ$ and
$\phi_{\rm sh} = 0^\circ$. Each curve corresponds
to the shower profile with fixed observer position $\phi$ while varying $\theta$
(see caption of Fig.~\ref{fig:inclined45}). Plotted for $\phi = 0^\circ, 30^\circ, 60^\circ$ and
$90^\circ$; profiles along $\phi$ and $\phi + 180^\circ$ are plotted using a single curve
with the latter corresponding to negative values of $\theta$. The shower leaves 
medium 1 (ice) into medium 2 (air) while crossing at shower maximum. The
electric field is evaluated at a frequency of 50~MHz (top) and 1~GHz (bottom). Each
curve is an average of a hundred showers.  }
\label{fig:cut45} 
\end{minipage}
\hfill 
\begin{minipage}[t]{.49\textwidth}
\includegraphics[width = \textwidth]{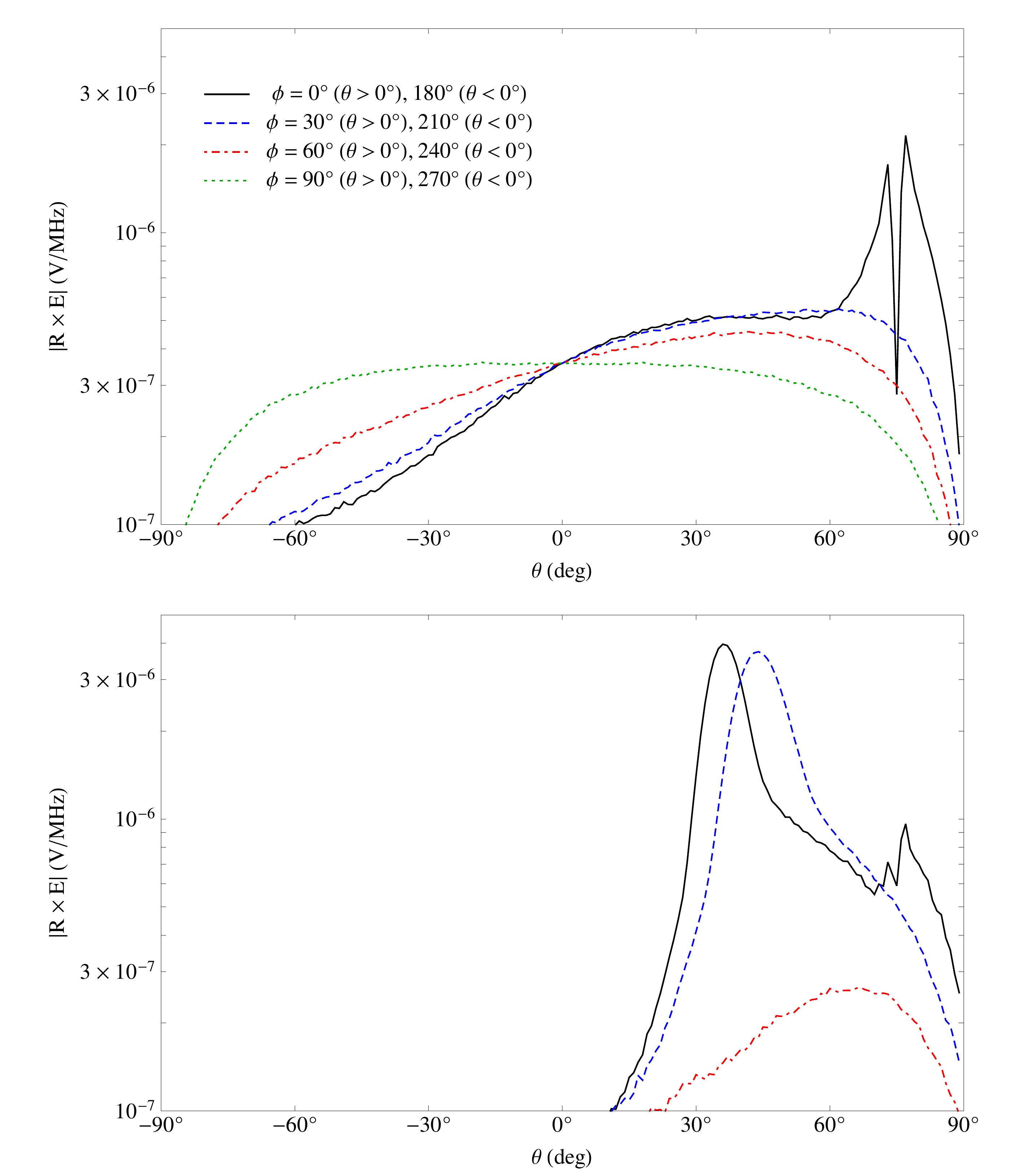}
\caption{{Same as Fig.~\ref{fig:cut45} for a shower with incident zenith angle of
$\theta_{\rm sh} = 75^\circ$ and $\phi_{\rm sh} = 0^\circ$.}}
\label{fig:cut75}
\vfill
\end{minipage}
\end{figure*}

To gain insight on the properties of TR from inclined showers, we performed simulations of
100~TeV showers with large zenith angle crossing the boundary from ice to air. The
shower's starting point in ice was chosen so that the shower maximum development would
occur at the intersection of the shower axis and the boundary plane. Results of the
simulations for two showers with incident angles of $\theta_{\rm sh} = 45^\circ$ and $75^\circ$,
respectively, are first shown in Figs.~\ref{fig:inclined45} and \ref{fig:inclined75}. In
these graphs, each point corresponds to the electric field in a direction $(\theta,\phi)$ of
observation, the polar angle of the plot being equal to $\phi$ and its radial distance
from the origin proportional to $\theta$. Due to symmetry of the situation, we show
only half of the sky in each plot. The direction of the shower axis $(\theta_{\rm
sh},\phi_{\rm sh} = 0^\circ)$ is marked by a cross. The electric field was evaluated at 50 and
1~GHz. In Figs.~\ref{fig:cut45} and \ref{fig:cut75} we show the electric field for the
same physical situation evaluated along lines of constant $\phi$ for $\phi = 0^\circ,
30^\circ, 60^\circ$ and $90^\circ$; profiles along $\phi$ and $\phi + 180^\circ$ are
plotted using a single curve with the latter corresponding to negative values of $\theta$.
Each curve is an average over a hundred showers.

Several distinctive features characterize the radiation. At 50~MHz, the electric field is
sizable over a large portion of the hemisphere even for the $75^\circ$ shower, with an
intensity comparable to that of a vertical shower. Also, there is a small angular region
around the direction of the shower axis where the electric field increases by a factor of
about three, clearly visible in Fig.~\ref{fig:cut45} and \ref{fig:cut75}. 
This peak corresponds to Cherenkov emission in air by high energy shower
particles moving close to the shower axis (see left panel of
Fig.~\ref{fig:energy_decomposition}). As also evident in Figs.~\ref{fig:cut45} and
\ref{fig:cut75}, the electric field drops
significantly for $(\theta,\phi)=(\theta_{\rm sh},\phi_{\rm sh})$ due to vanishing
$\boldsymbol{v}_\perp$ in Eq.~\eqref{ZHSFormula} for $\boldsymbol{v} \parallel
\boldsymbol{k}$.

At 1~GHz, the emission is overall more beamed towards the direction of the
shower axis, and this effect is more pronounced the higher the frequency and the incident
angle of the shower. For the $75^\circ$ shower, $|R \times {\bf E}|$ is greater than
$10^{-6.5}\, \mathrm{V/MHz}$ in $\sim50\%$ of the hemisphere at 50~MHz, and only in
$\sim20\%$ of the hemisphere at 1~GHz. Another distinctive feature at high frequency is
the appearance of an angular band with significantly increased signal. The band is clearly
evident at 1~GHz, where the electric field is an order of magnitude larger. Its origin is
explained in terms of coherent Cherenkov radiation refracted from ice into air.  Indeed,
we found that most of the radiation in this angular region is coming from below the
boundary (``refracted contribution'').  Further insight is given by a simple model where
the Cherenkov cone produced by a highly relativistic particle moving in ice along the
shower axis is refracted into air according to Snell's law. The refracted Cherenkov cone
matches very well the position and shape of the observed band, confirming its association
to coherent Cherenkov emission by shower particles in ice. 

At 10 MHz, the emitted radiation is basically indistinguishable for showers leaving
either ice or soil when both are crossing the interface at their shower maximum. At higher frequencies both
electric field profiles show rise of the Cherenkov peak and increased beaming towards the
direction of the shower axis; these features start to appear at lower frequencies in ice
for reasons explained in section \ref{vertical6}.
 
\section{Conclusions}
\label{sec:discussion}

We have developed a general method to calculate the electric field radiated by the
ensemble of particles of a high-energy electromagnetic shower developing in two different
media. The algorithm is based on the standard ZHS Monte Carlo approach, where a particle
track is divided into sub-tracks, each contributing to the total electric field. In the
special case of a particle crossing the two media, the particle track is split at the
boundary, and the electric field from the two sub-tracks is evaluated taking properly into
account reflection and refraction at the boundary plane. Transition radiation naturally
arises from this procedure, in addition to the coherent Cherenkov radiation produced by
the shower particles within each medium. 

Our ZHS-TR algorithm has general applicability and could be used, for example, to evaluate
the radiation from an Extensive Air Shower hitting ground, as seen by an observer placed
above (or below) the Earth surface. In this paper, we focused our studies on the radiation
emitted at radio frequencies by showers transitioning from a dense medium into air. This
configuration is relevant for a shower originated by a UHE neutrino interaction below the
Earth surface, escaping ground into the atmosphere. The ZHS Monte Carlo treats only
electromagnetic showers, which are appropriate for a charged-current interaction of a
high-energy electron neutrino. However, we expect our results not to change significantly
for hadronic showers (produced in charged or neutral-current interactions of neutrinos of
any flavor) since a large fraction of their energy is ultimately dissipated in
electromagnetic processes \cite{AlvarezMuniz:1998px}. 
    
Fundamental properties of the radiation were determined through Monte Carlo simulations
of vertical showers crossing from ice to air. We found that the emission is fairly
isotropic, with an electric field strength of $\sim 5 \times 10^{-11}$~V/MHz/m for a
100~TeV shower observed at 10~km distance. The electric field scales approximately
linearly with the number of particles crossing the boundary and remains coherent up to
$\sim 1$~GHz. At EeV energies the growth of the signal magnitude with shower
energy is slower than linear, due to the LPM effect. Similar results were obtained when using
salt, soil or regolith as a dense medium. We also studied the characteristics of radio
emission from inclined showers. The radiation pattern is quite broad, but becomes
more beamed towards the direction of the shower axis for larger incident angles 
or higher emission frequencies. The electric field strength is similar to that of a
vertical shower, but increases significantly at frequencies of $\sim1$~GHz in a limited
angular band, corresponding to a coherent Cherenkov cone refracted from ice into air.
Electric field for showers leaving soil show similar features, only their onset appears at
higher frequencies due to better coherence in this medium.

Amongst these results, the angular distribution and frequency spectrum of the radiation
are particularly interesting. Given the wide solid angle of the emission, a large aperture
experiment becomes feasible. Also, substantial radiation in the GHz range facilitates
detection, thanks to the low radio background and advantages in detector design at these
frequencies. These observations encourage further studies to evaluate the potential of a
large-aperture UHE neutrino experiment based on detection of transition radiation.

\smallskip{\em Acknowledgments.--}  
The authors would like to dedicate this paper to Prof. Raymond Protheroe
who sadly passed away in July 2015. Ray got JAM and EZ interested in this
topic while visiting Santiago de Compostela in spring 2007 as part of his
sabbatical leave. We also thank Justin Bray and Clancy James for discussions
on this topic over the last years and Roland Crocker for early work. Finally,
we would like to thank the anonymous referee for his many invaluable suggestions.

This work was supported in part by NSF grant PHY-1412261 and by the
Kavli Institute for Cosmological Physics at the University of Chicago through
grant  NSF  PHY-1125897  and  an  endowment  from  the  Kavli  Foundation  and
its founder Fred Kavli. 
JAM and EZ additionally thank Ministerio de Econom\'ia (FPA2012-39489), 
Consolider-Ingenio 2010 CPAN Programme (CSD2007-00042), Xunta de
Galicia (GRC2013-024), Feder Funds and CESGA
(Centro de Supercomputaci\'{o}n de Galicia) for computing resources.

\bibliography{TRpaper}

\begin{thebibliography}{40}%
\makeatletter
\providecommand \@ifxundefined [1]{%
 \@ifx{#1\undefined}
}%
\providecommand \@ifnum [1]{%
 \ifnum #1\expandafter \@firstoftwo
 \else \expandafter \@secondoftwo
 \fi
}%
\providecommand \@ifx [1]{%
 \ifx #1\expandafter \@firstoftwo
 \else \expandafter \@secondoftwo
 \fi
}%
\providecommand \natexlab [1]{#1}%
\providecommand \enquote  [1]{``#1''}%
\providecommand \bibnamefont  [1]{#1}%
\providecommand \bibfnamefont [1]{#1}%
\providecommand \citenamefont [1]{#1}%
\providecommand \href@noop [0]{\@secondoftwo}%
\providecommand \href [0]{\begingroup \@sanitize@url \@href}%
\providecommand \@href[1]{\@@startlink{#1}\@@href}%
\providecommand \@@href[1]{\endgroup#1\@@endlink}%
\providecommand \@sanitize@url [0]{\catcode `\\12\catcode `\$12\catcode
  `\&12\catcode `\#12\catcode `\^12\catcode `\_12\catcode `\%12\relax}%
\providecommand \@@startlink[1]{}%
\providecommand \@@endlink[0]{}%
\providecommand \url  [0]{\begingroup\@sanitize@url \@url }%
\providecommand \@url [1]{\endgroup\@href {#1}{\urlprefix }}%
\providecommand \urlprefix  [0]{URL }%
\providecommand \Eprint [0]{\href }%
\providecommand \doibase [0]{http://dx.doi.org/}%
\providecommand \selectlanguage [0]{\@gobble}%
\providecommand \bibinfo  [0]{\@secondoftwo}%
\providecommand \bibfield  [0]{\@secondoftwo}%
\providecommand \translation [1]{[#1]}%
\providecommand \BibitemOpen [0]{}%
\providecommand \bibitemStop [0]{}%
\providecommand \bibitemNoStop [0]{.\EOS\space}%
\providecommand \EOS [0]{\spacefactor3000\relax}%
\providecommand \BibitemShut  [1]{\csname bibitem#1\endcsname}%
\let\auto@bib@innerbib\@empty
\bibitem [{\citenamefont {Nagano}\ and\ \citenamefont
  {Watson}(2000)}]{Nagano:2000ve}%
  \BibitemOpen
  \bibfield  {author} {\bibinfo {author} {\bibfnamefont {M.}~\bibnamefont
  {Nagano}}\ and\ \bibinfo {author} {\bibfnamefont {A.~A.}\ \bibnamefont
  {Watson}},\ }\href {\doibase 10.1103/RevModPhys.72.689} {\bibfield  {journal}
  {\bibinfo  {journal} {Rev. Mod. Phys.}\ }\textbf {\bibinfo {volume} {72}},\
  \bibinfo {pages} {689} (\bibinfo {year} {2000})}\BibitemShut {NoStop}%
\bibitem [{\citenamefont {Watson}(2014)}]{Watson:2013cla}%
  \BibitemOpen
  \bibfield  {author} {\bibinfo {author} {\bibfnamefont {A.~A.}\ \bibnamefont
  {Watson}},\ }\href {\doibase 10.1088/0034-4885/77/3/036901} {\bibfield
  {journal} {\bibinfo  {journal} {Rept. Prog. Phys.}\ }\textbf {\bibinfo
  {volume} {77}},\ \bibinfo {pages} {036901} (\bibinfo {year} {2014})},\
  \Eprint {http://arxiv.org/abs/1310.0325} {arXiv:1310.0325 [astro-ph.HE]}
  \BibitemShut {NoStop}%
\bibitem [{\citenamefont {Abraham}\ \emph {et~al.}(2010)\citenamefont {Abraham}
  \emph {et~al.}}]{Abraham:2010mj}%
  \BibitemOpen
  \bibfield  {author} {\bibinfo {author} {\bibfnamefont {J.}~\bibnamefont
  {Abraham}} \emph {et~al.} (\bibinfo {collaboration} {Pierre Auger}),\ }\href
  {\doibase 10.1016/j.physletb.2010.02.013} {\bibfield  {journal} {\bibinfo
  {journal} {Phys.Lett.}\ }\textbf {\bibinfo {volume} {B685}},\ \bibinfo
  {pages} {239} (\bibinfo {year} {2010})},\ \Eprint
  {http://arxiv.org/abs/1002.1975} {arXiv:1002.1975 [astro-ph.HE]} \BibitemShut
  {NoStop}%
\bibitem [{\citenamefont {Fukushima}(2015)}]{Fukushima:2015bza}%
  \BibitemOpen
  \bibfield  {author} {\bibinfo {author} {\bibfnamefont {M.}~\bibnamefont
  {Fukushima}} (\bibinfo {collaboration} {Telescope Array}),\ }\href@noop {} {\
   (\bibinfo {year} {2015})},\ \Eprint {http://arxiv.org/abs/1503.06961}
  {arXiv:1503.06961 [astro-ph.HE]} \BibitemShut {NoStop}%
\bibitem [{\citenamefont {Aab}\ \emph {et~al.}(2015)\citenamefont {Aab} \emph
  {et~al.}}]{Aab:2015zoa}%
  \BibitemOpen
  \bibfield  {author} {\bibinfo {author} {\bibfnamefont {A.}~\bibnamefont
  {Aab}} \emph {et~al.} (\bibinfo {collaboration} {Pierre Auger}),\ }\href@noop
  {} {\bibfield  {journal} {\bibinfo  {journal} {Submitted to: Nucl. Instrum.
  Meth.}\ } (\bibinfo {year} {2015})},\ \Eprint
  {http://arxiv.org/abs/1502.01323} {arXiv:1502.01323 [astro-ph.IM]}
  \BibitemShut {NoStop}%
\bibitem [{\citenamefont {Abu-Zayyad}\ \emph {et~al.}(2012)\citenamefont
  {Abu-Zayyad} \emph {et~al.}}]{AbuZayyad:2012kk}%
  \BibitemOpen
  \bibfield  {author} {\bibinfo {author} {\bibfnamefont {T.}~\bibnamefont
  {Abu-Zayyad}} \emph {et~al.} (\bibinfo {collaboration} {Telescope Array}),\
  }\href {\doibase 10.1016/j.nima.2012.05.079} {\bibfield  {journal} {\bibinfo
  {journal} {Nucl. Instrum. Meth.}\ }\textbf {\bibinfo {volume} {A689}},\
  \bibinfo {pages} {87} (\bibinfo {year} {2012})},\ \Eprint
  {http://arxiv.org/abs/1201.4964} {arXiv:1201.4964 [astro-ph.IM]} \BibitemShut
  {NoStop}%
\bibitem [{\citenamefont {Kotera}\ and\ \citenamefont
  {Olinto}(2011)}]{Kotera:2011cp}%
  \BibitemOpen
  \bibfield  {author} {\bibinfo {author} {\bibfnamefont {K.}~\bibnamefont
  {Kotera}}\ and\ \bibinfo {author} {\bibfnamefont {A.~V.}\ \bibnamefont
  {Olinto}},\ }\href {\doibase 10.1146/annurev-astro-081710-102620} {\bibfield
  {journal} {\bibinfo  {journal} {Ann. Rev. Astron. Astrophys.}\ }\textbf
  {\bibinfo {volume} {49}},\ \bibinfo {pages} {119} (\bibinfo {year} {2011})},\
  \Eprint {http://arxiv.org/abs/1101.4256} {arXiv:1101.4256 [astro-ph.HE]}
  \BibitemShut {NoStop}%
\bibitem [{\citenamefont {Aartsen}\ \emph {et~al.}(2013)\citenamefont {Aartsen}
  \emph {et~al.}}]{Aartsen:2013jdh}%
  \BibitemOpen
  \bibfield  {author} {\bibinfo {author} {\bibfnamefont {M.}~\bibnamefont
  {Aartsen}} \emph {et~al.} (\bibinfo {collaboration} {IceCube}),\ }\href
  {\doibase 10.1126/science.1242856} {\bibfield  {journal} {\bibinfo  {journal}
  {Science}\ }\textbf {\bibinfo {volume} {342}},\ \bibinfo {pages} {1242856}
  (\bibinfo {year} {2013})},\ \Eprint {http://arxiv.org/abs/1311.5238}
  {arXiv:1311.5238 [astro-ph.HE]} \BibitemShut {NoStop}%
\bibitem [{\citenamefont {Kravchenko}\ \emph {et~al.}(2003)\citenamefont
  {Kravchenko} \emph {et~al.}}]{Kravchenko:2001id}%
  \BibitemOpen
  \bibfield  {author} {\bibinfo {author} {\bibfnamefont {I.}~\bibnamefont
  {Kravchenko}} \emph {et~al.} (\bibinfo {collaboration} {RICE}),\ }\href
  {\doibase 10.1016/S0927-6505(02)00194-9} {\bibfield  {journal} {\bibinfo
  {journal} {Astropart. Phys.}\ }\textbf {\bibinfo {volume} {19}},\ \bibinfo
  {pages} {15} (\bibinfo {year} {2003})},\ \Eprint
  {http://arxiv.org/abs/astro-ph/0112372} {arXiv:astro-ph/0112372 [astro-ph]}
  \BibitemShut {NoStop}%
\bibitem [{\citenamefont {Gorham}\ \emph {et~al.}(2009)\citenamefont {Gorham}
  \emph {et~al.}}]{Gorham:2008dv}%
  \BibitemOpen
  \bibfield  {author} {\bibinfo {author} {\bibfnamefont {P.~W.}\ \bibnamefont
  {Gorham}} \emph {et~al.} (\bibinfo {collaboration} {ANITA}),\ }\href
  {\doibase 10.1016/j.astropartphys.2009.05.003} {\bibfield  {journal}
  {\bibinfo  {journal} {Astropart. Phys.}\ }\textbf {\bibinfo {volume} {32}},\
  \bibinfo {pages} {10} (\bibinfo {year} {2009})},\ \Eprint
  {http://arxiv.org/abs/0812.1920} {arXiv:0812.1920 [astro-ph]} \BibitemShut
  {NoStop}%
\bibitem [{\citenamefont {Allison}\ \emph {et~al.}(2012)\citenamefont {Allison}
  \emph {et~al.}}]{Allison:2011wk}%
  \BibitemOpen
  \bibfield  {author} {\bibinfo {author} {\bibfnamefont {P.}~\bibnamefont
  {Allison}} \emph {et~al.},\ }\href {\doibase
  10.1016/j.astropartphys.2011.11.010} {\bibfield  {journal} {\bibinfo
  {journal} {Astropart. Phys.}\ }\textbf {\bibinfo {volume} {35}},\ \bibinfo
  {pages} {457} (\bibinfo {year} {2012})},\ \Eprint
  {http://arxiv.org/abs/1105.2854} {arXiv:1105.2854 [astro-ph.IM]} \BibitemShut
  {NoStop}%
\bibitem [{\citenamefont {Barwick}\ \emph {et~al.}(2015)\citenamefont {Barwick}
  \emph {et~al.}}]{Barwick:2014pca}%
  \BibitemOpen
  \bibfield  {author} {\bibinfo {author} {\bibfnamefont {S.~W.}\ \bibnamefont
  {Barwick}} \emph {et~al.} (\bibinfo {collaboration} {ARIANNA}),\ }\href
  {\doibase 10.1016/j.astropartphys.2015.04.002} {\bibfield  {journal}
  {\bibinfo  {journal} {Astropart. Phys.}\ }\textbf {\bibinfo {volume} {70}},\
  \bibinfo {pages} {12} (\bibinfo {year} {2015})},\ \Eprint
  {http://arxiv.org/abs/1410.7352} {arXiv:1410.7352 [astro-ph.HE]} \BibitemShut
  {NoStop}%
\bibitem [{\citenamefont {Wissel}\ \emph {et~al.}(2015)\citenamefont {Wissel}
  \emph {et~al.}}]{Wissel:2015aa}%
  \BibitemOpen
  \bibfield  {author} {\bibinfo {author} {\bibfnamefont {S.~A.}\ \bibnamefont
  {Wissel}} \emph {et~al.},\ }\href@noop {} {\bibfield  {journal} {\bibinfo
  {journal} {PoS(ICRC2015)}\ ,\ \bibinfo {pages} {1150}} (\bibinfo {year}
  {2015})}\BibitemShut {NoStop}%
\bibitem [{\citenamefont {James}\ \emph {et~al.}(2010)\citenamefont {James},
  \citenamefont {Ekers}, \citenamefont {Alvarez-Muniz}, \citenamefont {Bray},
  \citenamefont {McFadden}, \citenamefont {Phillips}, \citenamefont
  {Protheroe},\ and\ \citenamefont {Roberts}}]{James:2009sf}%
  \BibitemOpen
  \bibfield  {author} {\bibinfo {author} {\bibfnamefont {C.~W.}\ \bibnamefont
  {James}}, \bibinfo {author} {\bibfnamefont {R.~D.}\ \bibnamefont {Ekers}},
  \bibinfo {author} {\bibfnamefont {J.}~\bibnamefont {Alvarez-Muniz}}, \bibinfo
  {author} {\bibfnamefont {J.~D.}\ \bibnamefont {Bray}}, \bibinfo {author}
  {\bibfnamefont {R.~A.}\ \bibnamefont {McFadden}}, \bibinfo {author}
  {\bibfnamefont {C.~J.}\ \bibnamefont {Phillips}}, \bibinfo {author}
  {\bibfnamefont {R.~J.}\ \bibnamefont {Protheroe}}, \ and\ \bibinfo {author}
  {\bibfnamefont {P.}~\bibnamefont {Roberts}},\ }\href {\doibase
  10.1103/PhysRevD.81.042003} {\bibfield  {journal} {\bibinfo  {journal} {Phys.
  Rev.}\ }\textbf {\bibinfo {volume} {D81}},\ \bibinfo {pages} {042003}
  (\bibinfo {year} {2010})},\ \Eprint {http://arxiv.org/abs/0911.3009}
  {arXiv:0911.3009 [astro-ph.HE]} \BibitemShut {NoStop}%
\bibitem [{\citenamefont {Gorham}\ \emph {et~al.}(2004)\citenamefont {Gorham},
  \citenamefont {Hebert}, \citenamefont {Liewer}, \citenamefont {Naudet},
  \citenamefont {Saltzberg},\ and\ \citenamefont {Williams}}]{Gorham:2003da}%
  \BibitemOpen
  \bibfield  {author} {\bibinfo {author} {\bibfnamefont {P.~W.}\ \bibnamefont
  {Gorham}}, \bibinfo {author} {\bibfnamefont {C.~L.}\ \bibnamefont {Hebert}},
  \bibinfo {author} {\bibfnamefont {K.~M.}\ \bibnamefont {Liewer}}, \bibinfo
  {author} {\bibfnamefont {C.~J.}\ \bibnamefont {Naudet}}, \bibinfo {author}
  {\bibfnamefont {D.}~\bibnamefont {Saltzberg}}, \ and\ \bibinfo {author}
  {\bibfnamefont {D.}~\bibnamefont {Williams}},\ }\href {\doibase
  10.1103/PhysRevLett.93.041101} {\bibfield  {journal} {\bibinfo  {journal}
  {Phys. Rev. Lett.}\ }\textbf {\bibinfo {volume} {93}},\ \bibinfo {pages}
  {041101} (\bibinfo {year} {2004})},\ \Eprint
  {http://arxiv.org/abs/astro-ph/0310232} {arXiv:astro-ph/0310232 [astro-ph]}
  \BibitemShut {NoStop}%
\bibitem [{\citenamefont {Beresnyak}\ \emph {et~al.}(2005)\citenamefont
  {Beresnyak}, \citenamefont {Dagkesamansky}, \citenamefont {Kovalenko},
  \citenamefont {Oreshko},\ and\ \citenamefont
  {Zheleznykh}}]{Beresnyak:2005yh}%
  \BibitemOpen
  \bibfield  {author} {\bibinfo {author} {\bibfnamefont {A.~R.}\ \bibnamefont
  {Beresnyak}}, \bibinfo {author} {\bibfnamefont {R.~D.}\ \bibnamefont
  {Dagkesamansky}}, \bibinfo {author} {\bibfnamefont {A.~V.}\ \bibnamefont
  {Kovalenko}}, \bibinfo {author} {\bibfnamefont {V.~V.}\ \bibnamefont
  {Oreshko}}, \ and\ \bibinfo {author} {\bibfnamefont {I.~M.}\ \bibnamefont
  {Zheleznykh}},\ }\href {\doibase 10.1134/1.1862359} {\bibfield  {journal}
  {\bibinfo  {journal} {Astron. Rep.}\ }\textbf {\bibinfo {volume} {49}},\
  \bibinfo {pages} {127} (\bibinfo {year} {2005})}\BibitemShut {NoStop}%
\bibitem [{\citenamefont {Askar'yan}(1962)}]{Askaryan:1962oya}%
  \BibitemOpen
  \bibfield  {author} {\bibinfo {author} {\bibfnamefont {G.}~\bibnamefont
  {Askar'yan}},\ }\href@noop {} {\bibfield  {journal} {\bibinfo  {journal}
  {Sov.Phys.JETP}\ }\textbf {\bibinfo {volume} {14}},\ \bibinfo {pages} {441}
  (\bibinfo {year} {1962})}\BibitemShut {NoStop}%
\bibitem [{\citenamefont {Ginzburg}\ and\ \citenamefont
  {Frank}(1945)}]{Ginzburg:1945zz}%
  \BibitemOpen
  \bibfield  {author} {\bibinfo {author} {\bibfnamefont {V.}~\bibnamefont
  {Ginzburg}}\ and\ \bibinfo {author} {\bibfnamefont {I.}~\bibnamefont
  {Frank}},\ }\href@noop {} {\bibfield  {journal} {\bibinfo  {journal}
  {J.Phys.(USSR)}\ }\textbf {\bibinfo {volume} {9}},\ \bibinfo {pages} {353}
  (\bibinfo {year} {1945})}\BibitemShut {NoStop}%
\bibitem [{\citenamefont {Ter-Mikaelian}(1972)}]{Ter-Mikaelian:1972aa}%
  \BibitemOpen
  \bibfield  {author} {\bibinfo {author} {\bibfnamefont {M.~L.}\ \bibnamefont
  {Ter-Mikaelian}},\ }\href@noop {} {\emph {\bibinfo {title} {{High energy
  electromagnetic processes in condensed media}}}},\ Internat. Sci. Tracts
  Phys. Astron.\ (\bibinfo  {publisher} {Wiley},\ \bibinfo {address} {New York,
  NY},\ \bibinfo {year} {1972})\BibitemShut {NoStop}%
\bibitem [{\citenamefont {Gazazian}\ \emph {et~al.}(2001)\citenamefont
  {Gazazian}, \citenamefont {Ispirian},\ and\ \citenamefont
  {Vardanyan}}]{Gazazian:2001aa}%
  \BibitemOpen
  \bibfield  {author} {\bibinfo {author} {\bibfnamefont {E.~D.}\ \bibnamefont
  {Gazazian}}, \bibinfo {author} {\bibfnamefont {K.}~\bibnamefont {Ispirian}},
  \ and\ \bibinfo {author} {\bibfnamefont {A.~S.}\ \bibnamefont {Vardanyan}},\
  }\href@noop {} {\bibfield  {journal} {\bibinfo  {journal} {AIP Conf. Proc.}\
  }\textbf {\bibinfo {volume} {597}},\ \bibinfo {pages} {111} (\bibinfo {year}
  {2001})}\BibitemShut {NoStop}%
\bibitem [{\citenamefont {de~Vries}\ \emph {et~al.}(2015)\citenamefont
  {de~Vries}, \citenamefont {Buitink}, \citenamefont {van Eijndhoven},
  \citenamefont {Meures}, \citenamefont {O'Murchadha} \emph
  {et~al.}}]{deVries:2015oda}%
  \BibitemOpen
  \bibfield  {author} {\bibinfo {author} {\bibfnamefont {K.~D.}\ \bibnamefont
  {de~Vries}}, \bibinfo {author} {\bibfnamefont {S.}~\bibnamefont {Buitink}},
  \bibinfo {author} {\bibfnamefont {N.}~\bibnamefont {van Eijndhoven}},
  \bibinfo {author} {\bibfnamefont {T.}~\bibnamefont {Meures}}, \bibinfo
  {author} {\bibfnamefont {A.}~\bibnamefont {O'Murchadha}},  \emph {et~al.},\
  }\href@noop {} {\  (\bibinfo {year} {2015})},\ \Eprint
  {http://arxiv.org/abs/1503.02808} {arXiv:1503.02808 [astro-ph.HE]}
  \BibitemShut {NoStop}%
\bibitem [{\citenamefont {Falcke}\ \emph {et~al.}(2004)\citenamefont {Falcke},
  \citenamefont {Gorham},\ and\ \citenamefont {Protheroe}}]{Falcke:2004aw}%
  \BibitemOpen
  \bibfield  {author} {\bibinfo {author} {\bibfnamefont {H.}~\bibnamefont
  {Falcke}}, \bibinfo {author} {\bibfnamefont {P.}~\bibnamefont {Gorham}}, \
  and\ \bibinfo {author} {\bibfnamefont {R.~J.}\ \bibnamefont {Protheroe}},\
  }\bibfield  {booktitle} {\emph {\bibinfo {booktitle} {{International SKA
  Conference 2003 Geraldton, Australia, July 27-August 2, 2003}}},\ }\href
  {\doibase 10.1016/j.newar.2004.09.028} {\bibfield  {journal} {\bibinfo
  {journal} {New Astron. Rev.}\ }\textbf {\bibinfo {volume} {48}},\ \bibinfo
  {pages} {1487} (\bibinfo {year} {2004})},\ \Eprint
  {http://arxiv.org/abs/astro-ph/0409229} {arXiv:astro-ph/0409229 [astro-ph]}
  \BibitemShut {NoStop}%
\bibitem [{\citenamefont {Zas}\ \emph {et~al.}(1992)\citenamefont {Zas},
  \citenamefont {Halzen},\ and\ \citenamefont {Stanev}}]{Zas:1991jv}%
  \BibitemOpen
  \bibfield  {author} {\bibinfo {author} {\bibfnamefont {E.}~\bibnamefont
  {Zas}}, \bibinfo {author} {\bibfnamefont {F.}~\bibnamefont {Halzen}}, \ and\
  \bibinfo {author} {\bibfnamefont {T.}~\bibnamefont {Stanev}},\ }\href
  {\doibase 10.1103/PhysRevD.45.362} {\bibfield  {journal} {\bibinfo  {journal}
  {Phys.Rev.}\ }\textbf {\bibinfo {volume} {D45}},\ \bibinfo {pages} {362}
  (\bibinfo {year} {1992})}\BibitemShut {NoStop}%
\bibitem [{\citenamefont {James}\ \emph {et~al.}(2011)\citenamefont {James},
  \citenamefont {Falcke}, \citenamefont {Huege},\ and\ \citenamefont
  {Ludwig}}]{James:2010vm}%
  \BibitemOpen
  \bibfield  {author} {\bibinfo {author} {\bibfnamefont {C.~W.}\ \bibnamefont
  {James}}, \bibinfo {author} {\bibfnamefont {H.}~\bibnamefont {Falcke}},
  \bibinfo {author} {\bibfnamefont {T.}~\bibnamefont {Huege}}, \ and\ \bibinfo
  {author} {\bibfnamefont {M.}~\bibnamefont {Ludwig}},\ }\href {\doibase
  10.1103/PhysRevE.84.056602} {\bibfield  {journal} {\bibinfo  {journal}
  {Phys.Rev.}\ }\textbf {\bibinfo {volume} {E84}},\ \bibinfo {pages} {056602}
  (\bibinfo {year} {2011})},\ \Eprint {http://arxiv.org/abs/1007.4146}
  {arXiv:1007.4146 [physics.class-ph]} \BibitemShut {NoStop}%
\bibitem [{\citenamefont {{Ginzburg}}\ and\ \citenamefont
  {{Tsytovich}}(1990)}]{Ginzburg:1990aa}%
  \BibitemOpen
  \bibfield  {author} {\bibinfo {author} {\bibfnamefont {V.~L.}\ \bibnamefont
  {{Ginzburg}}}\ and\ \bibinfo {author} {\bibfnamefont {V.~N.}\ \bibnamefont
  {{Tsytovich}}},\ }\href@noop {} {\emph {\bibinfo {title} {{Transition
  radiation and transition scattering.}}}}\ (\bibinfo  {publisher} {Hilger},\
  \bibinfo {address} {Bristol, UK},\ \bibinfo {year} {1990})\BibitemShut
  {NoStop}%
\bibitem [{\citenamefont {Allan}(1971)}]{Allan:1967}%
  \BibitemOpen
  \bibfield  {author} {\bibinfo {author} {\bibfnamefont {H.}~\bibnamefont
  {Allan}},\ }\href@noop {} {\emph {\bibinfo {title} {{Radio Emission from
  Extensive Air Showers}}}}\ (\bibinfo  {publisher} {North-Holland},\ \bibinfo
  {address} {Amsterdam},\ \bibinfo {year} {1971})\ pp.\ \bibinfo {pages}
  {169--302}\BibitemShut {NoStop}%
\bibitem [{\citenamefont {Garcia-Fernandez}\ \emph {et~al.}(2013)\citenamefont
  {Garcia-Fernandez}, \citenamefont {Alvarez-Muniz}, \citenamefont {Carvalho},
  \citenamefont {Romero-Wolf},\ and\ \citenamefont
  {Zas}}]{GarciaFernandez:2012yz}%
  \BibitemOpen
  \bibfield  {author} {\bibinfo {author} {\bibfnamefont {D.}~\bibnamefont
  {Garcia-Fernandez}}, \bibinfo {author} {\bibfnamefont {J.}~\bibnamefont
  {Alvarez-Muniz}}, \bibinfo {author} {\bibfnamefont {W.~R.}\ \bibnamefont
  {Carvalho}}, \bibinfo {author} {\bibfnamefont {A.}~\bibnamefont
  {Romero-Wolf}}, \ and\ \bibinfo {author} {\bibfnamefont {E.}~\bibnamefont
  {Zas}},\ }\href {\doibase 10.1103/PhysRevD.87.023003} {\bibfield  {journal}
  {\bibinfo  {journal} {Phys.Rev.}\ }\textbf {\bibinfo {volume} {D87}},\
  \bibinfo {pages} {023003} (\bibinfo {year} {2013})},\ \Eprint
  {http://arxiv.org/abs/1210.1052} {arXiv:1210.1052 [astro-ph.HE]} \BibitemShut
  {NoStop}%
\bibitem [{\citenamefont {Alvarez-Muniz}\ \emph {et~al.}(2009)\citenamefont
  {Alvarez-Muniz}, \citenamefont {James}, \citenamefont {Protheroe},\ and\
  \citenamefont {Zas}}]{AlvarezMuniz:2009zzc}%
  \BibitemOpen
  \bibfield  {author} {\bibinfo {author} {\bibfnamefont {J.}~\bibnamefont
  {Alvarez-Muniz}}, \bibinfo {author} {\bibfnamefont {C.}~\bibnamefont
  {James}}, \bibinfo {author} {\bibfnamefont {R.}~\bibnamefont {Protheroe}}, \
  and\ \bibinfo {author} {\bibfnamefont {E.}~\bibnamefont {Zas}},\ }\href
  {\doibase 10.1016/j.astropartphys.2009.06.005} {\bibfield  {journal}
  {\bibinfo  {journal} {Astropart.Phys.}\ }\textbf {\bibinfo {volume} {32}},\
  \bibinfo {pages} {100} (\bibinfo {year} {2009})}\BibitemShut {NoStop}%
\bibitem [{\citenamefont {Halzen}\ \emph {et~al.}(1991)\citenamefont {Halzen},
  \citenamefont {Zas},\ and\ \citenamefont {Stanev}}]{Halzen:1990vt}%
  \BibitemOpen
  \bibfield  {author} {\bibinfo {author} {\bibfnamefont {F.}~\bibnamefont
  {Halzen}}, \bibinfo {author} {\bibfnamefont {E.}~\bibnamefont {Zas}}, \ and\
  \bibinfo {author} {\bibfnamefont {T.}~\bibnamefont {Stanev}},\ }\href
  {\doibase 10.1016/0370-2693(91)91920-Q} {\bibfield  {journal} {\bibinfo
  {journal} {Phys.Lett.}\ }\textbf {\bibinfo {volume} {B257}},\ \bibinfo
  {pages} {432} (\bibinfo {year} {1991})}\BibitemShut {NoStop}%
\bibitem [{\citenamefont {Alvarez-Muniz}\ \emph {et~al.}(2006)\citenamefont
  {Alvarez-Muniz}, \citenamefont {Marques}, \citenamefont {Vazquez},\ and\
  \citenamefont {Zas}}]{AlvarezMuniz:2005qb}%
  \BibitemOpen
  \bibfield  {author} {\bibinfo {author} {\bibfnamefont {J.}~\bibnamefont
  {Alvarez-Muniz}}, \bibinfo {author} {\bibfnamefont {E.}~\bibnamefont
  {Marques}}, \bibinfo {author} {\bibfnamefont {R.~A.}\ \bibnamefont
  {Vazquez}}, \ and\ \bibinfo {author} {\bibfnamefont {E.}~\bibnamefont
  {Zas}},\ }\href {\doibase 10.1103/PhysRevD.74.023007} {\bibfield  {journal}
  {\bibinfo  {journal} {Phys.Rev.}\ }\textbf {\bibinfo {volume} {D74}},\
  \bibinfo {pages} {023007} (\bibinfo {year} {2006})},\ \Eprint
  {http://arxiv.org/abs/astro-ph/0512337} {arXiv:astro-ph/0512337 [astro-ph]}
  \BibitemShut {NoStop}%
\bibitem [{\citenamefont {Alvarez-Muniz}\ \emph {et~al.}(2000)\citenamefont
  {Alvarez-Muniz}, \citenamefont {Vazquez},\ and\ \citenamefont
  {Zas}}]{AlvarezMuniz:2000fw}%
  \BibitemOpen
  \bibfield  {author} {\bibinfo {author} {\bibfnamefont {J.}~\bibnamefont
  {Alvarez-Muniz}}, \bibinfo {author} {\bibfnamefont {R.}~\bibnamefont
  {Vazquez}}, \ and\ \bibinfo {author} {\bibfnamefont {E.}~\bibnamefont
  {Zas}},\ }\href {\doibase 10.1103/PhysRevD.62.063001} {\bibfield  {journal}
  {\bibinfo  {journal} {Phys.Rev.}\ }\textbf {\bibinfo {volume} {D62}},\
  \bibinfo {pages} {063001} (\bibinfo {year} {2000})},\ \Eprint
  {http://arxiv.org/abs/astro-ph/0003315} {arXiv:astro-ph/0003315 [astro-ph]}
  \BibitemShut {NoStop}%
\bibitem [{Note1()}]{Note1}%
  \BibitemOpen
  \bibinfo {note} {Note we use a non-standard convention.}\BibitemShut {Stop}%
\bibitem [{\citenamefont {Jackson}(1999)}]{Jackson:1999aa}%
  \BibitemOpen
  \bibfield  {author} {\bibinfo {author} {\bibfnamefont {J.~D.}\ \bibnamefont
  {Jackson}},\ }\href {http://cdsweb.cern.ch/record/490457} {\emph {\bibinfo
  {title} {Classical electrodynamics}}},\ \bibinfo {edition} {3rd}\ ed.\
  (\bibinfo  {publisher} {Wiley},\ \bibinfo {address} {New York, {NY}},\
  \bibinfo {year} {1999})\BibitemShut {NoStop}%
\bibitem [{\citenamefont {Huege}\ and\ \citenamefont
  {Falcke}(2003)}]{Huege:2003up}%
  \BibitemOpen
  \bibfield  {author} {\bibinfo {author} {\bibfnamefont {T.}~\bibnamefont
  {Huege}}\ and\ \bibinfo {author} {\bibfnamefont {H.}~\bibnamefont {Falcke}},\
  }\href {\doibase 10.1051/0004-6361:20031422} {\bibfield  {journal} {\bibinfo
  {journal} {Astron. Astrophys.}\ }\textbf {\bibinfo {volume} {412}},\ \bibinfo
  {pages} {19} (\bibinfo {year} {2003})},\ \Eprint
  {http://arxiv.org/abs/astro-ph/0309622} {arXiv:astro-ph/0309622 [astro-ph]}
  \BibitemShut {NoStop}%
\bibitem [{\citenamefont {Motloch}\ \emph {et~al.}(2014)\citenamefont
  {Motloch}, \citenamefont {Hollon},\ and\ \citenamefont
  {Privitera}}]{Motloch:2013kva}%
  \BibitemOpen
  \bibfield  {author} {\bibinfo {author} {\bibfnamefont {P.}~\bibnamefont
  {Motloch}}, \bibinfo {author} {\bibfnamefont {N.}~\bibnamefont {Hollon}}, \
  and\ \bibinfo {author} {\bibfnamefont {P.}~\bibnamefont {Privitera}},\ }\href
  {\doibase 10.1016/j.astropartphys.2013.11.002} {\bibfield  {journal}
  {\bibinfo  {journal} {Astropart.Phys.}\ }\textbf {\bibinfo {volume} {54}},\
  \bibinfo {pages} {40} (\bibinfo {year} {2014})},\ \Eprint
  {http://arxiv.org/abs/1309.0561} {arXiv:1309.0561 [astro-ph.IM]} \BibitemShut
  {NoStop}%
\bibitem [{\citenamefont {Landau}\ and\ \citenamefont
  {Pomeranchuk}(1953{\natexlab{a}})}]{Landau:1953um}%
  \BibitemOpen
  \bibfield  {author} {\bibinfo {author} {\bibfnamefont {L.~D.}\ \bibnamefont
  {Landau}}\ and\ \bibinfo {author} {\bibfnamefont {I.}~\bibnamefont
  {Pomeranchuk}},\ }\href@noop {} {\bibfield  {journal} {\bibinfo  {journal}
  {Dokl. Akad. Nauk Ser. Fiz.}\ }\textbf {\bibinfo {volume} {92}},\ \bibinfo
  {pages} {535} (\bibinfo {year} {1953}{\natexlab{a}})}\BibitemShut {NoStop}%
\bibitem [{\citenamefont {Landau}\ and\ \citenamefont
  {Pomeranchuk}(1953{\natexlab{b}})}]{Landau:1953gr}%
  \BibitemOpen
  \bibfield  {author} {\bibinfo {author} {\bibfnamefont {L.~D.}\ \bibnamefont
  {Landau}}\ and\ \bibinfo {author} {\bibfnamefont {I.}~\bibnamefont
  {Pomeranchuk}},\ }\href@noop {} {\bibfield  {journal} {\bibinfo  {journal}
  {Dokl. Akad. Nauk Ser. Fiz.}\ }\textbf {\bibinfo {volume} {92}},\ \bibinfo
  {pages} {735} (\bibinfo {year} {1953}{\natexlab{b}})}\BibitemShut {NoStop}%
\bibitem [{\citenamefont {Alvarez-Muniz}\ and\ \citenamefont
  {Zas}(1997)}]{AlvarezMuniz:1997sh}%
  \BibitemOpen
  \bibfield  {author} {\bibinfo {author} {\bibfnamefont {J.}~\bibnamefont
  {Alvarez-Muniz}}\ and\ \bibinfo {author} {\bibfnamefont {E.}~\bibnamefont
  {Zas}},\ }\href {\doibase 10.1016/S0370-2693(97)01009-5} {\bibfield
  {journal} {\bibinfo  {journal} {Phys. Lett.}\ }\textbf {\bibinfo {volume}
  {B411}},\ \bibinfo {pages} {218} (\bibinfo {year} {1997})},\ \Eprint
  {http://arxiv.org/abs/astro-ph/9706064} {arXiv:astro-ph/9706064 [astro-ph]}
  \BibitemShut {NoStop}%
\bibitem [{\citenamefont {Gandhi}\ \emph {et~al.}(1998)\citenamefont {Gandhi},
  \citenamefont {Quigg}, \citenamefont {Reno},\ and\ \citenamefont
  {Sarcevic}}]{Gandhi:1998ri}%
  \BibitemOpen
  \bibfield  {author} {\bibinfo {author} {\bibfnamefont {R.}~\bibnamefont
  {Gandhi}}, \bibinfo {author} {\bibfnamefont {C.}~\bibnamefont {Quigg}},
  \bibinfo {author} {\bibfnamefont {M.~H.}\ \bibnamefont {Reno}}, \ and\
  \bibinfo {author} {\bibfnamefont {I.}~\bibnamefont {Sarcevic}},\ }\href
  {\doibase 10.1103/PhysRevD.58.093009} {\bibfield  {journal} {\bibinfo
  {journal} {Phys.Rev.}\ }\textbf {\bibinfo {volume} {D58}},\ \bibinfo {pages}
  {093009} (\bibinfo {year} {1998})},\ \Eprint
  {http://arxiv.org/abs/hep-ph/9807264} {arXiv:hep-ph/9807264 [hep-ph]}
  \BibitemShut {NoStop}%
\bibitem [{\citenamefont {Alvarez-Muniz}\ and\ \citenamefont
  {Zas}(1998)}]{AlvarezMuniz:1998px}%
  \BibitemOpen
  \bibfield  {author} {\bibinfo {author} {\bibfnamefont {J.}~\bibnamefont
  {Alvarez-Muniz}}\ and\ \bibinfo {author} {\bibfnamefont {E.}~\bibnamefont
  {Zas}},\ }\href {\doibase 10.1016/S0370-2693(98)00905-8} {\bibfield
  {journal} {\bibinfo  {journal} {Phys. Lett.}\ }\textbf {\bibinfo {volume}
  {B434}},\ \bibinfo {pages} {396} (\bibinfo {year} {1998})},\ \Eprint
  {http://arxiv.org/abs/astro-ph/9806098} {arXiv:astro-ph/9806098 [astro-ph]}
  \BibitemShut {NoStop}%
\end{thebibliography}%

\end{document}